\documentclass[aps,prb,twocolumn,groupedaddress,showpacs,draft,intlimits,amsmath,amssymb,floats]{revtex4}

\usepackage{bm}
\usepackage[final]{graphicx}
\usepackage{epsfig}
\usepackage{subfigure}

\begin{document}

\title{Numerical exploration of spontaneous broken symmetries in multi-orbital Hubbard models} 

\author{Y. F. Kung$^{1,2}$}
\thanks{The authors contributed equally to this work.}
\author{C.-C. Chen$^{3}$}
\thanks{The authors contributed equally to this work.}
\author{B. Moritz$^{2,4}$}
\author{S. Johnston$^{5,6}$}
\author{R. Thomale$^7$}
\author{T. P. Devereaux$^{2,8}$}

\affiliation{$^1$Department of Physics, Stanford University, Stanford, California 94305, USA}
\affiliation{$^2$Stanford Institute for Materials and Energy Sciences, SLAC National Accelerator Laboratory, Menlo Park, California 94025, USA}
\affiliation{$^3$Advanced Photon Source, Argonne National Laboratory, Lemont, Illinois 60439, USA}
\affiliation{$^4$Department of Physics and Astrophysics, University of North Dakota, Grand Forks, North Dakota 58202, USA}
\affiliation{$^5$Department of Physics and Astronomy, University of Tennessee, Knoxville, Tennessee 37996, USA}
\affiliation{$^6$Joint Institute for Advanced Materials, University of Tennessee, Knoxville, Tennessee 37996, USA}
\affiliation{$^7$Institute for Theoretical Physics, University of W\"{u}rzburg, D-97074 W\"{u}rzburg, Germany}
\affiliation{$^8$Geballe Laboratory for Advanced Materials, Stanford University, Stanford, California 94305, USA}

\begin{abstract}
We study three proposals for broken symmetry in the cuprate pseudogap -- oxygen antiferromagnetism, $\Theta_{II}$ orbital loop currents, and circulating currents involving apex oxygens -- through numerical exploration of multi-orbital Hubbard models.  Our numerically exact results show no evidence for the existence of oxygen antiferromagnetic order or the $\Theta_{II}$ phase in the three-orbital Hubbard model.
The model also fails to sustain an ordered current pattern even with the presence of additional apex oxygen orbitals.
We thereby conclude that it is difficult to stabilize the aforementioned phases in the multi-orbital Hubbard models for parameters relevant to cuprate superconductors. 
However, the $\Theta_{II}$ phase might be stabilized through explicit flux terms.  
We find an enhanced propensity for circulating currents with such terms in calculations simulating applied stress or strain, which skew the copper-oxygen plane to resemble a kagome lattice.
We propose an experimental viewpoint to shed additional light on this problem.
\end{abstract}


\maketitle

\section{Introduction}

One intriguing open question in cuprate superconductors concerns the nature of the enigmatic pseudogap regime,\cite{Timusk_RPP_1999,Norman_2005,Fradkin_arXiv_2014,Hashimoto_NatPhys_2014} which may be key to understanding the mechanism of high-temperature superconductivity.
A variety of theories have been proposed, falling into two broad categories.
The first describes the pseudogap as a region of preformed Cooper pairs that lack phase coherence above $T_c$, but cross over to the superconducting state below it.\cite{Emery_Nature_1995}
The second proposes that the pseudogap arises from a competing order, where possible ground states include antiferromagnetic (AFM) spin order on oxygen [Fig. 1(a)],\cite{Bourges_Comptes_2011} $\Theta_{II}$ orbital loop currents [Fig. 1(b)],\cite{Varma_PRB_1997, Varma_PRL_1999,Varma_PRB_2006, Varma_PRB_2012} currents that circulate between the planar and apex oxygens [Fig. 1(c)],\cite{Weber_PRL_2009} $d$-density waves,\cite{Chakravarty_PRB_2001,Chakravarty_PRB_2013} and charge density waves or nematic orders.\cite{Kivelson_RMP_2003,Lawler_Nature_2010, Nie_PNAS_2014}

To unravel this mystery, various experimental techniques have been employed.  
Probes of electronic structure, such as x-ray diffraction, resonant inelastic x-ray scattering (RIXS), angle-resolved photoemission spectroscopy (ARPES), scanning tunneling microscopy, and nuclear magnetic resonance (NMR) have uncovered evidence for a static or fluctuating charge density wave in the pseudogap regime.\cite{Chang_NatPhys_2012, Ghiringhelli_Science_2012,Hashimoto_NatPhys_2010, Ma_PRL_2008, Vershinin_Science_2004, Wise_NatPhys_2008,Hashimoto_PRB_2014,Vishik_PNAS_2012,Julien_Nature_2011,Julien_NatComm_2013,Julien_arXiv_2014}  
A magneto-optic Kerr effect study has reported time-reversal symmetry breaking\cite{Jia_PRL_2008} that follows a similar temperature dependence as pseudogap-related features from ARPES and optical spectroscopy.\cite{He_Science_2011}
Here the sign of the Kerr signal cannot be trained by magnetic fields, suggesting a connection to striped phases\cite{Kapitulnik_PRL_2012} or chiral charge orders.\cite{Hosur_PRB_2013, Orenstein_PRB_2013, Sebastian_Nature_2014, Hovnatan_PRL_2014}
Neutron scattering studies of YBa$_2$Cu$_3$O$_{6+x}$, HgBa$_2$CuO$_{4+\delta}$, and Bi$_2$Sr$_2$CaCu$_2$O$_{8+\delta}$ have seen evidence of intra-unit-cell magnetic order\cite{Bourges_Comptes_2011,Li_Nature_2008,Li_Nature_2010,Sidis_arXiv_2013} with an out-of-plane moment $\sim0.1 \mu_B$.\cite{Fauque_PRL_2006}
This order breaks time-reversal symmetry but preserves translational lattice symmetry with a $\mathbf{q}=0$ ordering vector.

These experiments have provoked great interest in the $\Theta_{II}$ phase because it is compatible with broken time-reversal symmetry (seen also by one ARPES study with circularly polarized photons\cite{Kaminski_Nature_2002}) while preserving translational symmetry, as observed in neutron scattering.
However, while accompanied by several distinct experimental features,\cite{Simon_PRL_2002, Berg_PRL_2008,Allais_PRB_2012,Nielsen_PRB_2012,Kivelson_arXiv_2012,Wang_PRB_2013, Aji_PRB_2013} the circulating current phase is still highly disputed, as no experiments can directly probe such an order parameter.  In addition, the $\Theta_{II}$ phase (a $\mathbf{q}=0$ Fermi surface instability) faces the challenge of explaining the excitation gap in the pseudogap regime.
Intense effort has been devoted to searching indirectly for orbital loop currents, using techniques such as NMR\cite{Strassle_PRL_2011, Lederer_PRB_2012,Mounce_PRL_2013} and muon spin rotation,\cite{Sonier_PRL_2009, Huang_PRB_2012,Storchak_arXiv_2014} but their existence in the cuprates remains inconclusive.

To complement the experimental efforts, various simulations have been performed to search for $\Theta_{II}$ currents but lead to contradictory results.
Variational Monte Carlo studies have found orbital currents in multi-orbital Hubbard models that include axial orbitals involving the apical oxygen, copper $d_{3z^2-r^2}$, or copper $4s$ orbitals.\cite{Weber_PRL_2009,Weber_PRL_2014}
A bosonization study of a copper-oxygen ladder finds orbital currents over a large region of the phase diagram.\cite{Giamarchi_Bosonization_2008}  Self-consistent mean field theory also finds the $\Theta_{II}$ orbital current in the three-orbital Hubbard model in a certain parameter range, although the interaction strengths required to stabilize the phase are too large for the cuprates.\cite{Fischer_MFT_2011}  
However, an exact diagonalization study of the three-orbital $t-J$ model\cite{Thomale_PRL_2007,Thomale_PRB_2008} fails to find such a phase as the ground state. 
This is consistent with the results of a variational cluster approximation study \cite{Lu_VCA_2012}
and a density matrix renormalization group study of a copper-oxygen ladder.\cite{Scalapino_DMRG_2009}
The debate continues, though,
as the physics of a ladder may not represent that in the copper-oxygen plane, and the down-folded $t-J$ Hamiltonian cannot properly account for charge fluctuations and the multi-orbital nature of the problem.\cite{Weber_PRL_2009}

Here we investigate this issue by studying the multi-orbital Hubbard model using exact diagonalization (ED) and numerically exact determinant quantum Monte Carlo (DQMC) on small clusters.  
These techniques have the advantages of treating the two-dimensional system as opposed to ladders, as well as the full multi-band model as opposed to downfolded models, enabling us to shed new light on the question.
We focus on the scenarios suggested by neutron scattering experiments: oxygen antiferromagnetism, $\Theta_{II}$ orbital loop currents, and circulating currents that involve the apex oxygens.
From the results of this study, we conclude that it is unlikely for these phases to be stabilized as the ground state for  parameters relevant to cuprate superconductors.\cite{Hybertsen_PRB_1989}
On the other hand, the circulating current phase can be stabilized on a skewed kagome lattice when flux terms are included in the Hamiltonian.
We thereby propose experiments with stress or strain\cite{Chu_Science_2012, Hovnatan_PRL_2014} applied along the oxygen-oxygen bonds to elucidate the problem.

The paper is organized as follows:
Section II introduces the three-orbital model and the numerical techniques employed to solve the Hamiltonian.
Section III discusses magnetic ordering on copper and oxygen, as determined by spin-spin correlation functions.
Section IV focuses on the $\Theta_{II}$ phase and current-current correlations.
Results obtained with extensions of the three-orbital model are discussed in Sections V and VI, for calculations with an apex oxygen and with staggered flux applied to a skewed kagome lattice, respectively.
Section VII summarizes our findings with additional remarks.

\begin{figure}[t!]
\includegraphics[width=\columnwidth]{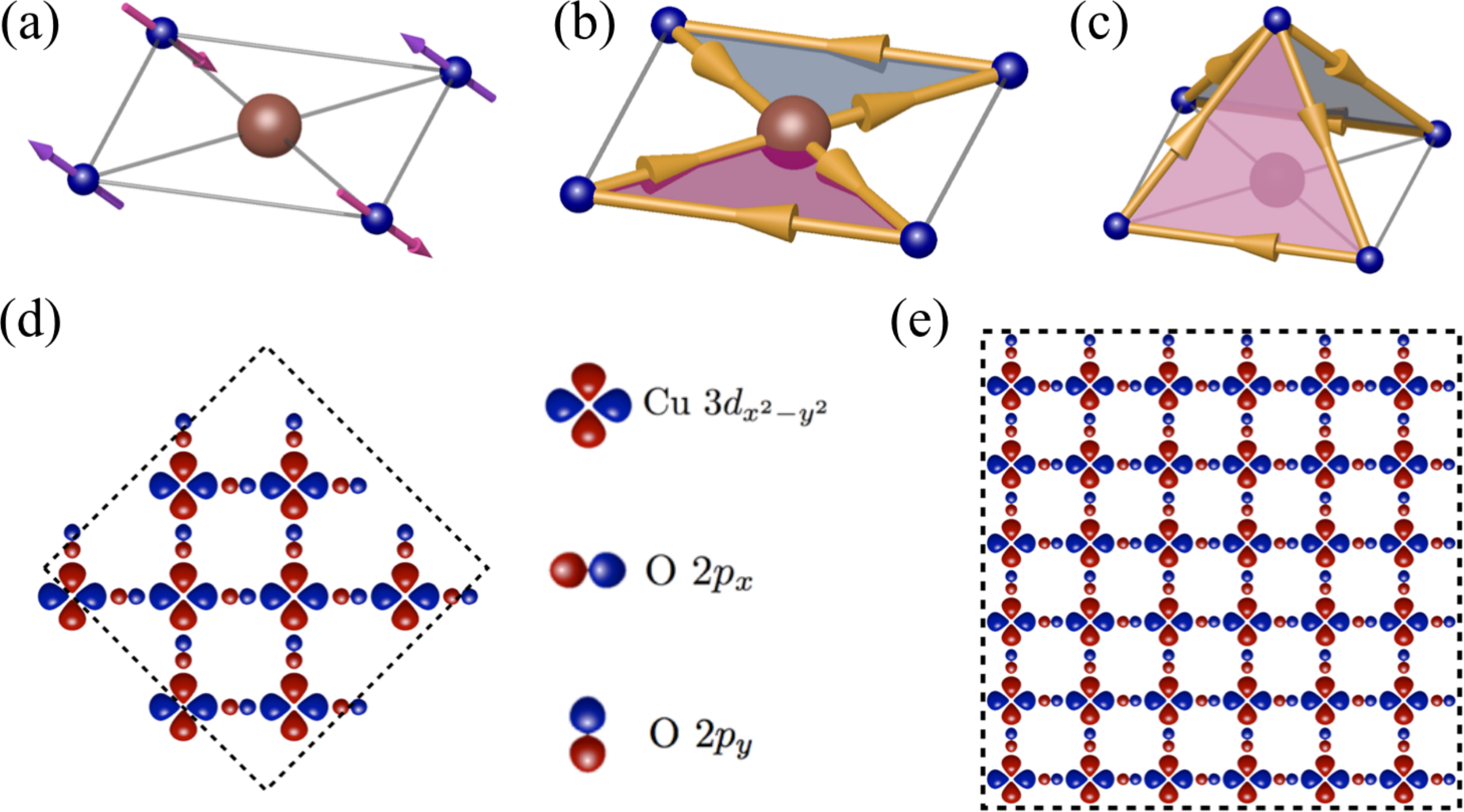}
\caption{
Cartoons of three proposed scenarios for the pseudogap: (a) oxygen antiferromagnetism, (b) $\Theta_{II}$ orbital loop currents, and (c) circulating currents involving apex oxygens.
The copper (blue) spheres represent the copper (oxygen) atoms.
(d) Cu$_8$O$_{16}$ cluster used in exact diagonalization. (e) Cu$_{36}$O$_{72}$ cluster used in determinant quantum Monte Carlo.  The calculations are performed with periodic boundary conditions.
}
\end{figure}

\section{Model and numerical methods} 
Most of the relevant physics in the cuprates is believed to occur in the copper-oxide planes;\cite{Dagotto_RMP_1994} as a starting point, we consider the three-orbital Hubbard model:\cite{Mattheiss_PRL_1987,Varma_SSC_1987, Emery_PRL_1987}
\begin{eqnarray}
H&=&-t_{pd}\sum_{\langle i,j \rangle \sigma} (d^\dagger_{i,\sigma} c_{j,\sigma}^{\phantom\dagger} + h.c.)\nonumber\\
&&-t_{pp}\sum_{\langle j,j^{'} \rangle \sigma} (c_{j,\sigma}^\dagger c_{j^{'},\sigma} + h.c.)\nonumber\\
&&+(\epsilon_d-\mu)\sum_{i,\sigma}n^d_{i,\sigma}+(\epsilon_p-\mu)\sum_{j,\sigma}n^p_{j,\sigma}\nonumber\\
&&+U_{dd}\sum_i n^d_{i,\uparrow}n^d_{i,\downarrow}+U_{pp}\sum_j n^p_{j,\uparrow}n^p_{j,\downarrow},
\end{eqnarray}\label{eq:H}
where $d^\dagger_{i,\sigma}$ ($d_{i,\sigma}^{\phantom\dagger}$) creates (destroys) a hole with spin $\sigma$ on a copper orbital at site $i$ and $c_{j,\sigma}^\dagger$ ($c_{j,\sigma}^{\phantom\dagger}$) creates (destroys) a hole with spin $\sigma$ on an oxygen orbital at site $j$.  
The first term in the Hamiltonian describes the hopping between copper 3$d_{x^2-y^2}$ and oxygen 2$p_{x,y}$ orbitals with an energy governed by $t_{pd}$.  
The second term describes the hopping between oxygens with an energy controlled by $t_{pp}$.   
$\epsilon_d$ and $\epsilon_p$ are the site energies on the copper 3$d_{x^2-y^2}$ and oxygen 2$p_{x,y}$ orbitals, and $U_{dd}$ and $U_{pp}$ parametrize the copper and oxygen on-site interactions.  
For simplicity, the unit cell length is set as $a=1$.

Unless noted otherwise, the parameters used in the simulations are (in units of eV) $U_{dd}=10.5$, $U_{pp}=4$, $t_{pd}=1.5$, $t_{pp}=0.65$, $\epsilon_d=0$, and $\epsilon_p=3.6$.\cite{Hybertsen_PRB_1989}  
Note, however, that our conclusions remain the same with different parameter sets proposed for the cuprates. \cite{Ohta_PRB_1991,Johnston_EPL_2009,Czyzyk_PRB_1994}
The effect of changes to these parameters will be examined in this manuscript.
The model is studied at $0\%$ and $12.5\%$ hole doping by the complementary, numerically exact techniques of DQMC and ED, where the filling is set by the chemical potential $\mu$ in the former and by the particle number sector (e.g. 5 spin up and 4 spin down holes for $12.5\%$ hole doping) in the latter. 
The ED calculations also include an inter-site interaction term, $V_{pd}\sum_{<ij>\sigma\sigma'} n^d_{i,\sigma} n^p_{j,\sigma'}$, with $V_{pd} = 1.2$. 

ED is a wave-function-based technique performed on finite-size clusters at zero temperature.\cite{Dagotto_RMP_1994}
This study uses Cu$_8$O$_{16}$ [Fig. 1(d)] and Cu$_8$O$_{24}$ (including additional apex oxygens) clusters with periodic boundary conditions.
The Hamiltonian matrices are constructed in a basis of momentum eigenstates,
and the resulting matrix eigenvalue problem is solved by iterative Krylov subspace methods.\cite{ARPACK}
DQMC is an imaginary-time, auxiliary-field method that computes observables from Green's functions.\cite{BSS_PRD_1981,Dopf_PRB_1990,Scalettar_PRB_1991,Dopf_PRL_1992}  It accesses larger system sizes (Cu$_{16}$O$_{32}$ and Cu$_{36}$O$_{72}$ [Fig. 1(e)]) but suffers from the fermion sign problem.  Hence DQMC simulations are performed at higher temperatures, making it complementary to ED.

These calculations have been performed on the largest system sizes possible given the limitations of Hilbert space dimensions in ED and fermion sign problems in DQMC.  At the current time, significantly larger system sizes in ED and lower temperatures in DQMC are not feasible.  Although these numerical techniques cannot access the thermodynamic limit, previous work has demonstrated that they reliably reproduce a number of experimental features, including Zhang-Rice singlets, RIXS excitations, Raman measurements, and ARPES spectra.\cite{Moritz_NJOP_2009,Chen_PRL_2010,Moritz_PRB_2011,Jia_NJOP_2012,Chen_PRB_2013,Jia_NatComm_2014}  These earlier results give us confidence in the numerically exact techniques.

\section{Oxygen antiferromagnetism}

We first examine the scenario of intra-unit-cell oxygen antiferromagnetic (AFM) order [Fig. 1(a)] by studying the real-space spin-spin correlation functions $S^{zz}(i, j) \equiv \langle S^z_i S^z_j\rangle - \langle S^z_i \rangle  \langle S^z_j \rangle$.
We note that on general grounds, spins in a two-dimensional Heisenberg system do not order at temperature $T\neq 0$. The spin correlation length remains finite but does grow exponentially with decreasing temperature as $\exp(C/T)$.
 Therefore, even on finite-size clusters, it is possible to identify long-range ordered phases from the behavior of the correlations as a function of cluster size or temperature.

\begin{figure}[t!]
\includegraphics[width=\columnwidth]{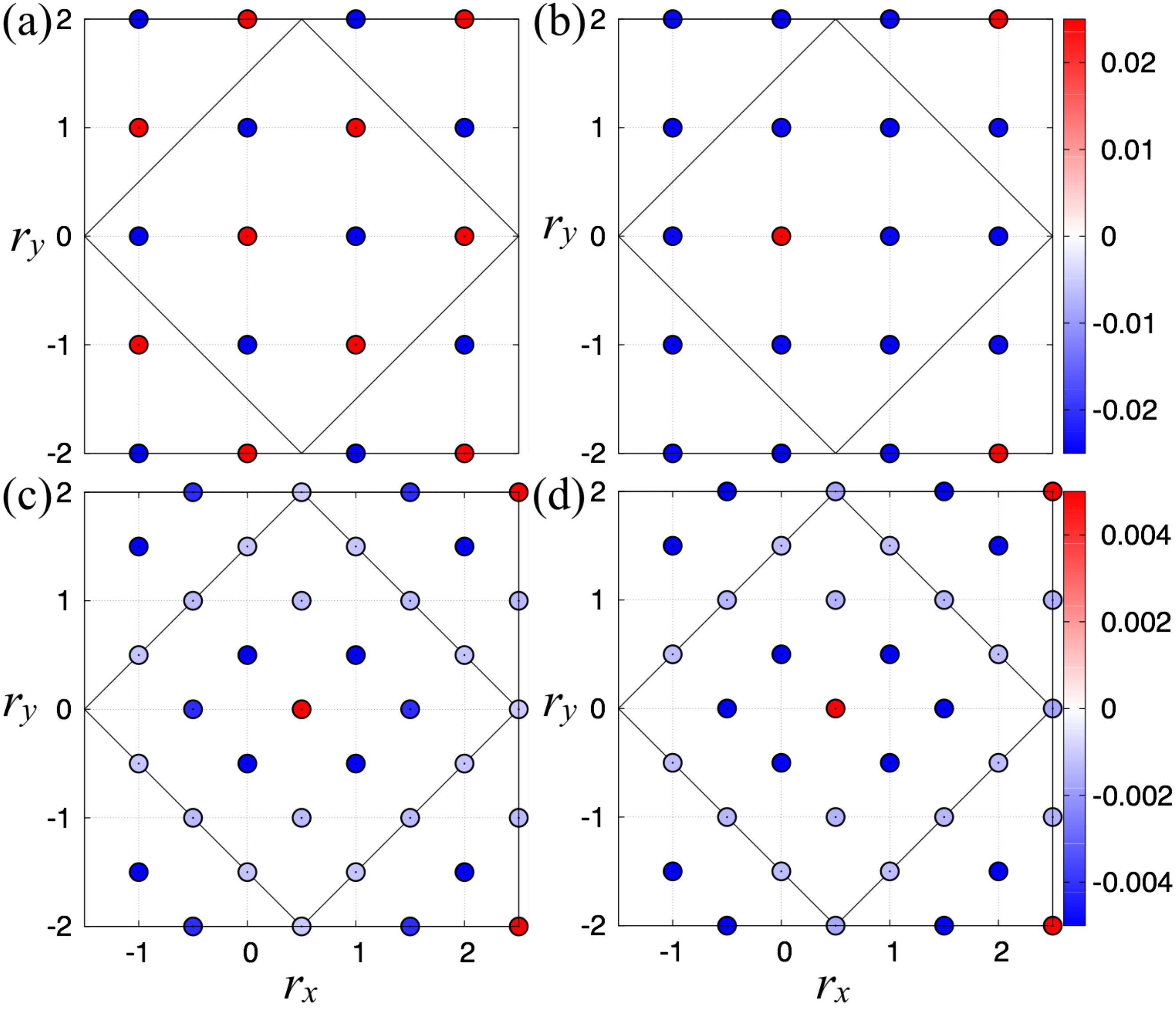}
\caption{
Spin-spin correlation functions calculated by ED on copper [(a)-(b)] and oxygen [(c)-(d)]. 
At $0\%$ doping [panel (a)], the copper spins show $(\pi,\pi)$ AFM correlations, which are suppressed at $12.5\%$ hole doping [panel (b)]. 
On the other hand, the oxygen spins do not exhibit AFM correlations at either doping level.
The box indicates the Cu$_8$O$_{16}$ cluster.
Values are referenced to the Cu $3d_{x^2-y^2}$ and O $2p_x$ orbitals in the $\mathbf{r}\equiv (r_x,r_y)=(0,0)$ unit cell for copper and oxygen spin-spin correlations, respectively.
}
\end{figure}

\begin{figure}[t!]
\includegraphics[width=\columnwidth]{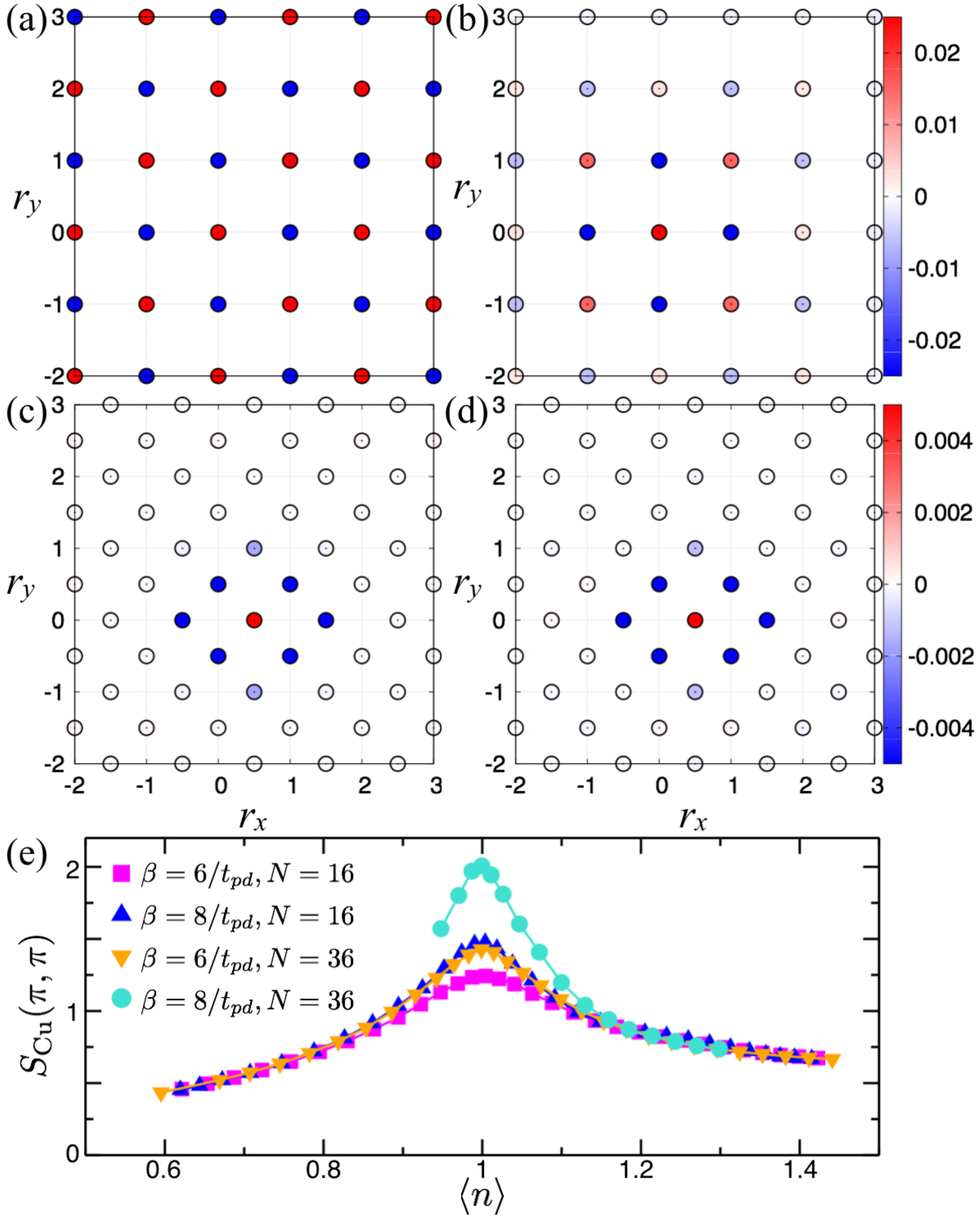}
\caption{
Spin-spin correlation functions on copper [(a)-(b)] and oxygen [(c)-(d)] from DQMC simulations at an inverse temperature $\beta=8/t_{pd}$.
At $0\%$ doping [panel (a)], the copper spins show $(\pi,\pi)$ AFM correlations, which are suppressed at $12.5\%$ hole doping [panel (b)]. 
In contrast, the oxygen spins do not exhibit AFM correlations at either doping level.
Values are referenced to the Cu $3d_{x^2-y^2}$ and O $2p_x$ orbitals in the $\mathbf{r}\equiv (r_x,r_y)=(0,0)$ unit cell for copper and oxygen spin-spin correlations, respectively.
(e) The copper $(\pi,\pi)$ spin structure factor. It peaks increasingly sharply at half filling with increasing $N$ (number of unit cells) or decreasing temperature. $\langle n \rangle$ is the number of holes per unit cell.
}
\end{figure}

Figure 2 shows the spin-spin correlation functions calculated using ED for the undoped and $12.5\%$ hole-doped cases.
As Fig. 2(a) demonstrates, the undoped system exhibits clear long-range $(\pi,\pi)$ AFM correlations on copper.
The order is suppressed upon hole doping [Fig. 2(b)], which is a well-known trend in the cuprate phase diagram.
In contrast, on the oxygen orbitals there is no sign of intra-unit-cell AFM correlations either undoped [Fig. 2(c)] or for $12.5\%$ hole-doping [Fig. 2(d)].
Note that in both cases, the magnitudes of the oxygen-oxygen spin-spin correlations are at least an order of magnitude smaller than those on copper.

The spin-spin correlation functions also have been calculated using DQMC at finite temperatures ($\beta\equiv 1/k_BT=6/t_{pd}$ and $\beta=8/t_{pd}$, on the order of 1000K) on larger Cu$_{16}$O$_{32}$ and Cu$_{36}$O$_{72}$ clusters.
As in ED, the undoped system shows clear long-range $(\pi,\pi)$ AFM correlations on copper [Fig. 3(a)],
which weaken with a reduced correlation length at $12.5\%$ hole doping [Fig. 3(b)].
There are no apparent AFM correlations on oxygen, as shown in Figs. 3(c) and (d), and the correlations are at least an order of magnitude smaller than those on copper.
With DQMC we have confirmed that $(\pi,\pi)$ represents the dominant spin ordering wave vector on copper at half filling.
Although the peak in the structure factor is suppressed rapidly upon either hole or electron doping [Fig. 3(e)], it peaks increasingly sharply with increasing system size $N$ or decreasing temperature, showing the expected trend toward N\'{e}el order at half filling.
This demonstrates the capability of DQMC to identify low-temperature ordered phases even with simulations performed at higher temperatures.\cite{White_PRB_1989}

It is evident from Figs. 2 and 3 that spins do not develop intra-unit-cell AFM order on oxygen, and the correlations would be too weak to explain the experimentally observed magnetic moment of $\sim 0.1\mu_B$.
Thus, our calculations do not support oxygen antiferromagnetism as a viable explanation for the cuprate pseudogap.

\section{$\Theta_{II}$ orbital loop currents}

We next study an explanation for the pseudogap based on the $\Theta_{II}$ orbital loop currents,
where circulating currents spontaneously develop and flow in a pattern similar to that depicted in Fig. 1(b).
Such a phase breaks the $C_4$ rotational and time-reversal symmetries, but preserves the product symmetry, as well as translational symmetry.
If present, a tendency towards this symmetry breaking can be identified from the current-current correlation functions $\langle j_{kl}(\mathbf{r}) j_{mn}(\mathbf{r'})  \rangle$,
where the oxygen-oxygen and copper-oxygen current operators at site $i$ are defined as:
\begin{eqnarray}
j_{\alpha\alpha'}&=&\frac{i t_{pp}}{\hbar}\sum_\sigma (c^{\alpha\dag}_{i\sigma} c^{\alpha '}_{j\sigma}-c^{\alpha ' \dag}_{j\sigma} c^{\alpha}_{i\sigma}),\nonumber\\
j_{\alpha d}&=&\frac{i t_{pd}}{\hbar}\sum_\sigma (d^\dag_{i\sigma} c^{\alpha}_{j\sigma}-c^{\alpha\dag}_{j\sigma} d_{i\sigma}^{\phantom\dagger}),\nonumber\\
\end{eqnarray}
where $\alpha$ indicates a $p_x$ or $p_y$ orbital.
The correlation functions are referenced to an oxygen-oxygen current in the $\mathbf{r}\equiv (r_x,r_y)=(2,2)$ and $(0,0)$ unit cells for ED and DQMC, respectively.  

\begin{figure}[t!]
\includegraphics[width=\columnwidth]{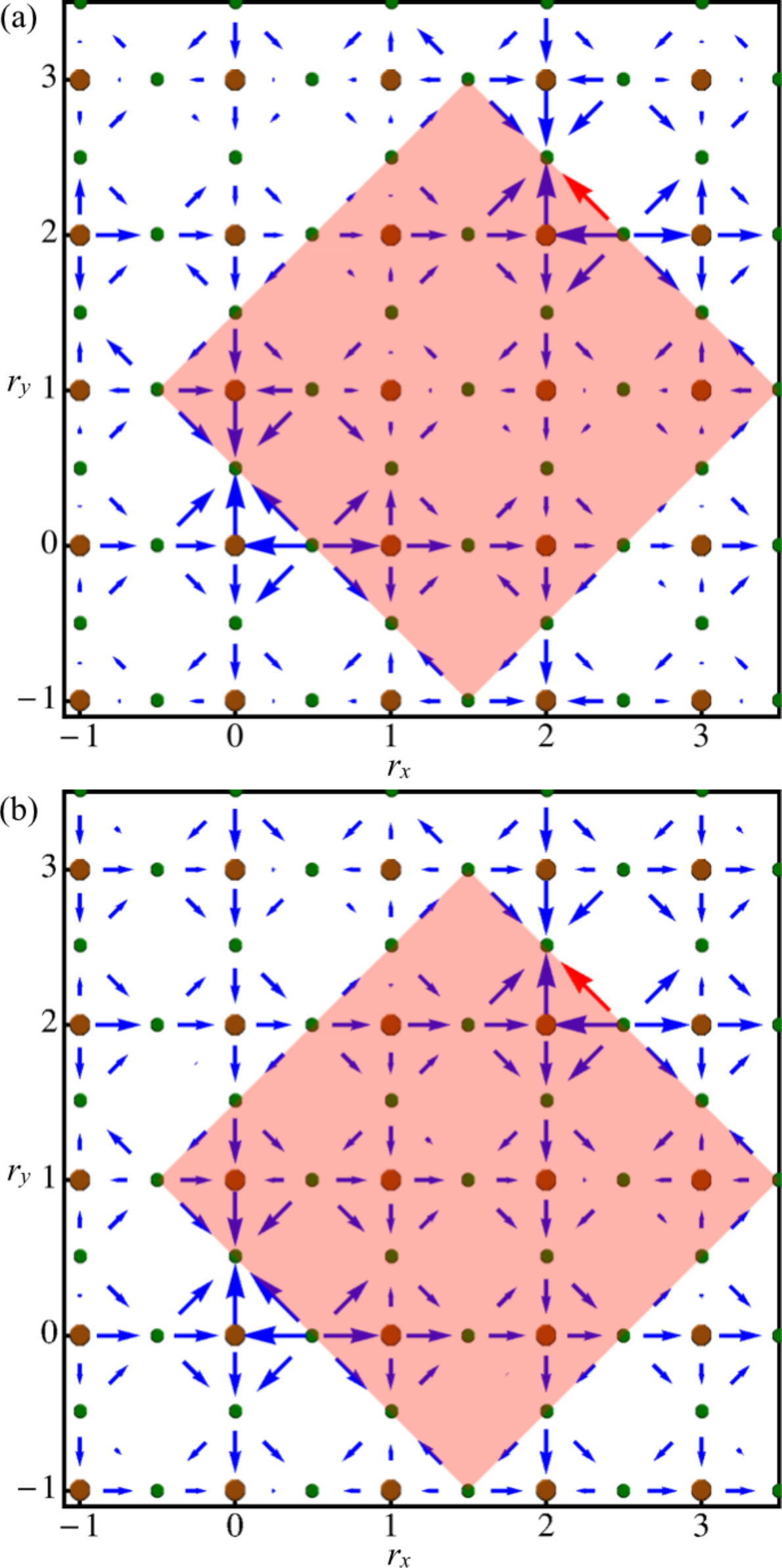}
\caption{
Current-current correlations calculated by ED at $12.5\%$ hole doping for (a) $\Delta=3.6$ eV and (b) $\Delta=0$.
The arrow length represents the correlation strength on a logarithmic scale, and the arrow head shows the current direction.
The reference oxygen-oxygen current is shown as a red arrow, whose length is 9.364 $(\textrm{eV}/\hbar)^2$ and 16.398 $(\textrm{eV}/\hbar)^2$
in (a) and (b), respectively.
The pink square indicates the Cu$_8$O$_{16}$ cluster, and the copper (green) circles represent the copper (oxygen) atoms.
From a comparison of (a) and (b), the correlations are enhanced when $\Delta$ decreases, but the currents do not exhibit the $\Theta_{II}$ pattern.
}
\end{figure}

Figure 4(a) shows the current-current correlation functions obtained from ED.
The magnitudes decrease rapidly away from the reference current, and the results show no sign of particular orbital loop current patterns.
This agrees qualitatively and semi-quantitatively with previous ED studies of the three-orbital $t-J$ model.\cite{Thomale_PRL_2007}
To compare to experiments, an upper bound on the magnetic moment is calculated from the correlations $ \langle j^2 \rangle < 5\times10^{-4} (\textrm{eV}/\hbar)^2$,
obtained using the oxygen currents farthest from the reference link.
With a copper-oxygen bond length $\sim 1.9$ {\AA}, the magnetic moment induced by the currents is found to be at most $\sim$0.025 $\mu_B$ (including contributions from both triangular current loops) per unit cell.  This number is approximately four times smaller than that derived from experiments, indicating that in this parameter regime, the current-induced moments would be too weak to explain the neutron scattering results.\cite{Fauque_PRL_2006}

A systematic exploration of parameters to enhance the circulating currents can be guided by mean-field theory. \cite{Fischer_MFT_2011}
In particular, a smaller charge-transfer gap $\Delta\equiv \epsilon_p - \epsilon_d$ has been suggested to favor the $\Theta_{II}$ phase.
We have performed ED calculations with $\Delta=0$, shifting more holes from copper to oxygen.
Indeed, most current-current correlations increase from those obtained with $\Delta=3.6$ eV from $\sim 50\%$ up to one order of magnitude [Fig. 4(b)].

However, the requirement that the undoped system be a charge-transfer insulator with an indirect band gap $\sim 2$ eV strongly constrains the value of $\Delta$.  With $\Delta=3.6$ eV in ED calculations, the energy gap between the lowest unoccupied peak at $(\pi,0)$ and the highest occupied state at $(\pi/2,\pi/2)$ is $\sim2$ eV, \cite{Chen_PRB_2013} in agreement with experiments.
The DQMC results are similar, but with the spectral features broadened and the gap reduced to $\sim1.5$ eV, due to finite temperatures. 
Other computational techniques have drawn similar conclusions, including a dynamical mean field theory study that determined $\Delta$ to be on the order of a few eV in order for the optical gap to fall in the experimentally observed range.\cite{Millis_PRB_2011}
Therefore, although a decreased $\Delta$ strengthens the current-current correlations, the calculated gap with $\Delta=0$ ($<0.3$ eV) is much smaller than that experimentally observed, ruling out the possibility of a vanishingly small $\Delta$.

\begin{figure}[t!]
\includegraphics[width=\columnwidth]{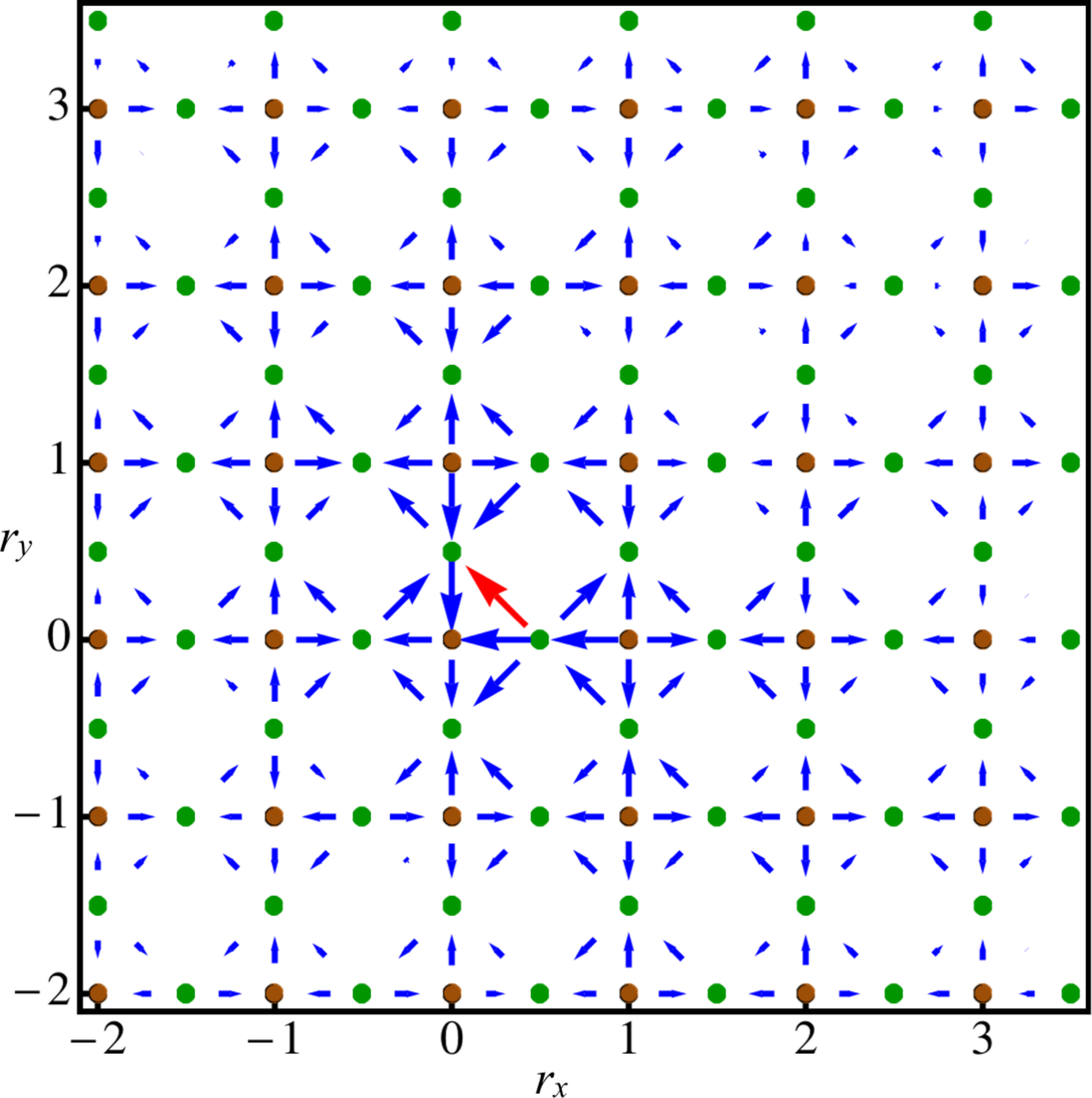}
\caption{
Current-current correlations at $12.5\%$ hole doping computed by DQMC on Cu$_{36}$O$_{72}$.
The arrow length represents the correlation strength on a logarithmic scale, and the arrow head shows the current direction.
The reference oxygen-oxygen current is shown as a red arrow, whose length is 4.875 $(\textrm{eV}/\hbar)^2$.
The copper (green) circles represent the copper (oxygen) atoms.
}
\end{figure}

\begin{table}[b!]
\caption{Summary of how parameters affect the current-current correlations.  $\uparrow$ indicates that the parameter enhances the correlation strengths, $\downarrow$ indicates that it destabilizes them, and -- indicates that it does not impact them significantly.}
\centering
\begin{tabular*}{\hsize}{@{\extracolsep{\fill} } c c c c c c c c}
\hline\hline
Method & $\Delta$ & $U_{pp}$ & $V_{pd}$ & $V_{pp}$ & $t_{pp}'$  &$\beta$ & $N$ \\ 
\hline
ED & $\downarrow$ & $\downarrow$ & $\uparrow$ & $\uparrow$ & $\uparrow$ & N/A & N/A \\
DQMC & $\downarrow$ & $\downarrow$ & N/A & N/A & $\uparrow$ & -- & $\downarrow$ \\ 
\hline
\end{tabular*}
\label{table:nonlin}
\end{table}

In addition to $\Delta$, we study the impact of on-site oxygen repulsion $U_{pp}$, as well as inter-site interactions $V_{pd}$ and $V_{pp}$.
Also in agreement with mean-field theory, \cite{Fischer_MFT_2011} our ED study finds a slight enhancement of most  correlations when $U_{pp}=0$, demonstrating that a strong $U_{pp}$ could hinder the formation of circulating currents. When $U_{pp}=0$, the local correlation strength can be increased by $\sim 5 -10\%$, and the correlations one unit cell away along the diagonal direction are more strongly enhanced ($\sim20-100\%$).
In contrast to $U_{pp}$, $V_{pd}$ and $V_{pp}$ slightly increase the relevant current-current correlations, but not to an extent that would qualitatively alter our conclusions.  In addition, we explore the impact of next-nearest neighbor oxygen-oxygen hopping $t_{pp}'$, which has been proposed to be necessary for orbital loop order.\cite{Weber_PRL_2014}  Setting $t_{pp}' \sim 2 t_{pp} \sim t_{pd}$ helps circulating currents develop locally but does not stabilize long-range order.

To study the effect of larger system sizes, we perform DQMC calculations on the Cu$_{16}$O$_{32}$ and Cu$_{36}$O$_{72}$ clusters.
In agreement with the Cu$_8$O$_{16}$ ED results, correlations are suppressed rapidly upon moving away from the reference current.
In fact, they decrease nearly to zero just one unit cell away along the diagonal.
Moreover, the directions do not show an appropriately ordered pattern [Fig. 5], particularly that for the $\Theta_{II}$ phase.
Quantitatively, the correlation strengths decrease up to $\sim20\%$ as the system grows from Cu$_{16}$O$_{32}$ to Cu$_{36}$O$_{72}$.

\begin{figure}[t!]
\includegraphics[width=\columnwidth]{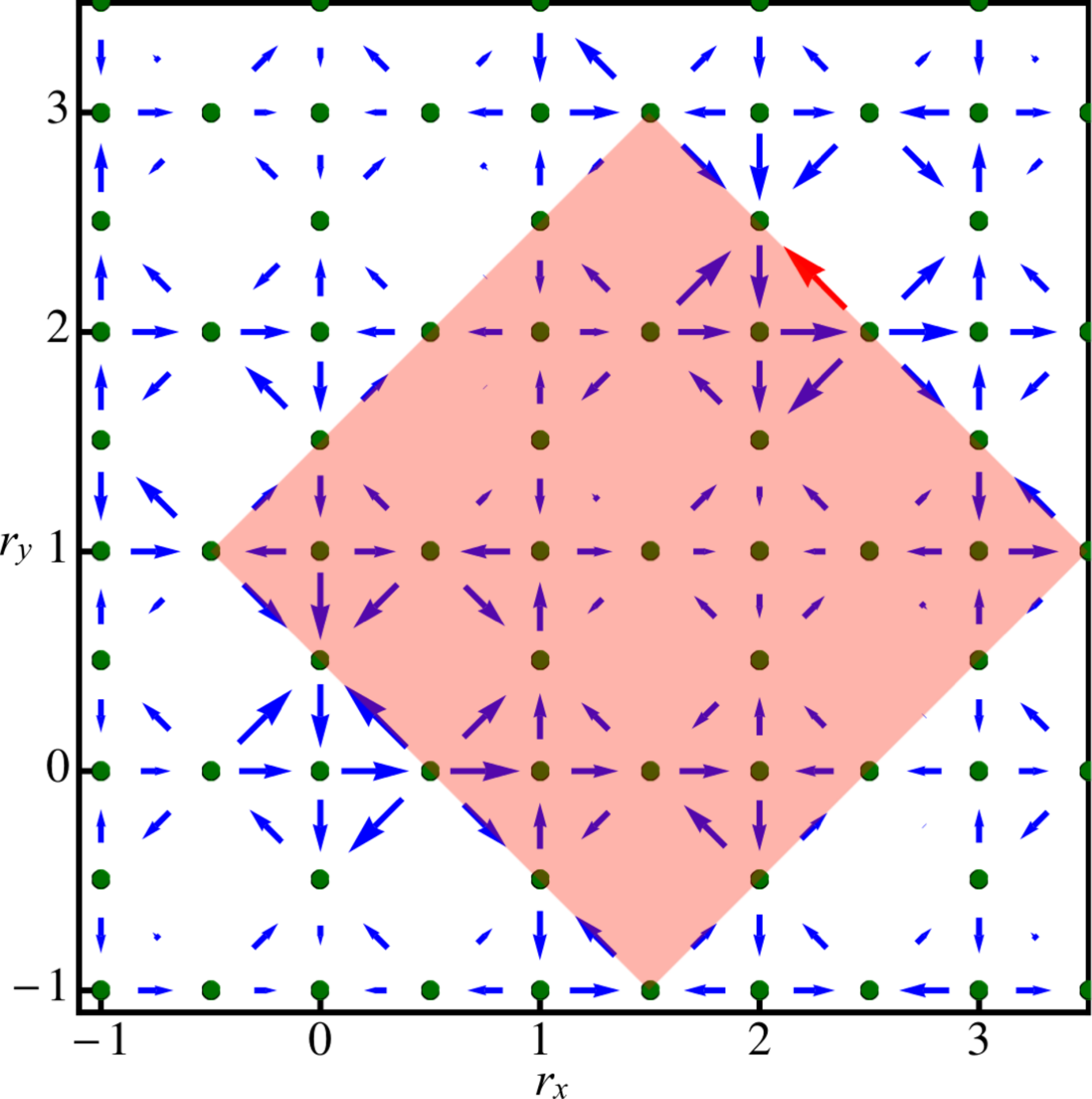}
\caption{
Current-current correlations at $12.5\%$ hole doping computed by ED on Cu$_{8}$O$_{24}$, which includes additional apex oxygens with $\epsilon_{p_z}=0$.
The arrow length represents the correlation strength on a logarithmic scale, and the arrow head shows the current direction.
The reference oxygen-oxygen current is shown as a red arrow, whose length is 13.918 $(\textrm{eV}/\hbar)^2$.
The pink square indicates the Cu$_8$O$_{24}$ cluster, and the green circles represent the oxygen atoms.
}
\end{figure}

In DQMC, the current-current correlations are enhanced as $\Delta$ decreases, in agreement with ED.
However, the increase is less significant, ranging from $\sim 50-100\%$, which results from thermal fluctuations weakening any orbital loop order. 
We also compare the correlations at $U_{pp}=0$ and $U_{pp}=4$ eV and determine that a significant decrease in the oxygen on-site repulsion leads to only a few percent increase in most current-current correlations.
Given this observation, we set $U_{pp}=0$ to access lower temperatures.  
Here the correlations are only weakly temperature-dependent,
suggesting that the $\Theta_{II}$ phase may not develop even in lower-temperature simulations.
Finally, simulations where $t_{pp}'\sim t_{pd}$ again show that next-nearest-neighbor oxygen-oxygen hopping enhances the circulating current pattern locally but cannot stabilize long-range order.

Note that in both ED and DQMC, varying the parameters affects the magnitudes of the correlations, but never changes the current directions to those expected in the $\Theta_{II}$ phase, regardless of $\Delta$, $U_{pp}$, $V_{pd}$, $V_{pp}$, $t_{pp}^{'}$, temperature, and system size $N$ (see Table I for a summary). From these results, we conclude that for parameters relevant to the cuprates, the three-orbital Hubbard model does not support a spontaneous $\Theta_{II}$ loop current ground state.

\section{Circulating currents involving apical oxygens}

To stabilize the circulating current phase, it may be necessary to extend the three-orbital model.
Polarized neutron scattering experiments have found a magnetic moment whose direction is tilted with respect to the crystal $c$-axis.
This could be explained by currents looping around the planar and apex oxygens [Fig. 1(c)].
To test this scenario, we extend our ED calculations by using a Cu$_8$O$_{24}$ cluster that includes additional apex oxygen $2p_z$ orbitals.
With an apex oxygen site energy $\epsilon_{p_z} = 3.6$ eV, we find that the current-current correlations between the planar and apex oxygens are two orders of magnitude smaller than the correlations between just the planar oxygens.
Therefore, the moment associated with the currents looping around the apical sites would be too weak to account for the experimentally observed moment of 0.1 $\mu_B$.
In addition, there is no sign of an ordered pattern.

Although we use $\epsilon_{p_z}=\epsilon_p$ in the simulation, the apical oxygen hole density is only $\sim0.3\%$, whereas the planar oxygen hole density is $\sim30\%$. This significant difference is mainly due to the additional hybridization pathway between the $2p_{x,y}$ and $3d_{x^2-y^2}$ orbitals.
On the other hand, the apical oxygen hole density can be stabilized by the copper $3d_{3z^2-r^2}$ orbital. \cite{Chen_PRL_2010}
Rather than increase the model's complexity by adding yet another orbital, this effect can be included by using a smaller apex oxygen site energy.
Figure 6 shows the current-current correlations computed with  $\epsilon_{p_z}=0$ and $\epsilon_p=3.6$ eV, where a substantial number of holes are transferred onto the apical sites. 
While the correlations are enhanced compared to the calculation with $\epsilon_{p_z}=\epsilon_p=3.6$ eV,
there is no circulating current pattern as depicted in Fig. 1(c), indicating that the presence of apex oxygen orbitals is not enough to stabilize the loop current phase.

\section{Flux applied to a skewed kagome lattice}

\begin{figure}[t!]
\includegraphics[width=\columnwidth]{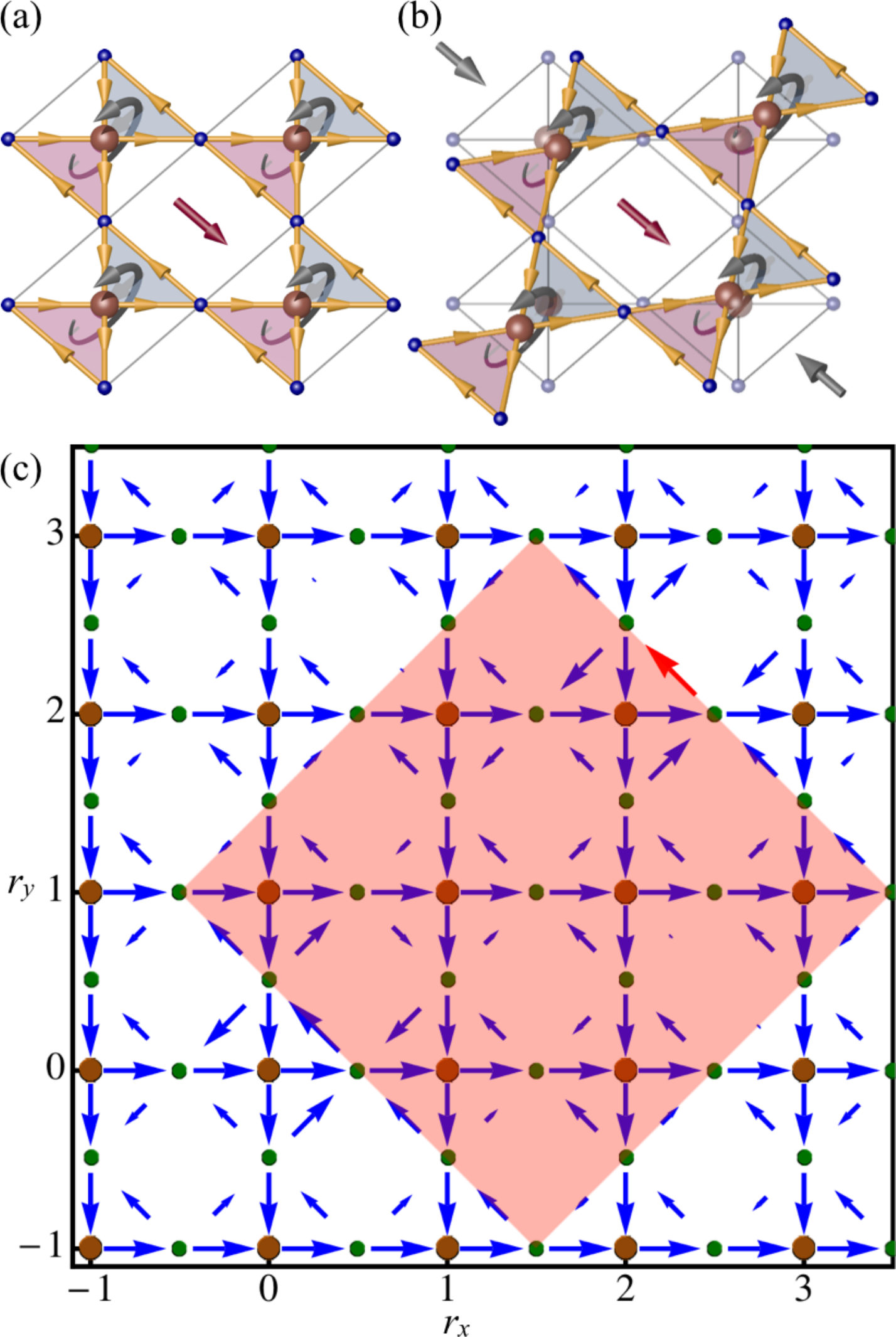}
\caption{
(a) A cartoon showing the $\Theta_{II}$ phase, whose current flow pattern assumes a skewed kagome lattice consisting of corner-sharing triangles.
The red arrow indicates the anapole moment direction.
(b) A schematic diagram showing stress or strain applied along the ($\bar{1}$10) direction,
which could enhance the orbital loop current strength if the order parameter is initially nonzero in the undistorted lattice.
(c) Current-current correlations calculated by ED with staggered flux terms at 12.5\% hole doping.
The arrow length represents the correlation strength on a logarithmic scale, and the arrow head shows the current direction.
The reference oxygen-oxygen current is shown as a red arrow, whose length is 15.580 $(\textrm{eV}/\hbar)^2$.
The pink square indicates the Cu$_8$O$_{16}$ cluster, and the copper (green) circles represent the copper (oxygen) atoms.
}
\end{figure}

In a final attempt to stabilize the $\Theta_{II}$ orbital loop currents, we add explicit staggered flux terms to the Hamiltonian.
Specifically, we replace the hybridization terms of the original three-orbital model by the time-reversal symmetry-breaking Hamiltonian $H'$:
\begin{eqnarray}
H'=&-&\sum_{\langle i,j \rangle \sigma}  t^{pd}_{ij}e^{i\phi_{pd}}(d^\dagger_{i,\sigma} c_{j,\sigma}^{\phantom\dagger} + h.c.)\nonumber
\\
&-&\sum_{\langle j,j^{'} \rangle \sigma}t^{pp}_{jj^{'}} e^{i\phi_{pp}} (c_{j,\sigma}^\dagger c_{j^{'},\sigma} + h.c.),
\end{eqnarray}
where the sign and distribution of the phases $\phi_{pd}$ and $\phi_{pp}$ are chosen to correspond to the current pattern in Fig. 7(a).
We then compute the current-current correlations with ED by varying the strength of the phases.
Figure 7(c) shows the results with $\phi_{pd}=\phi_{pp}$ and $\Delta \phi \equiv 2\phi_{pd} + \phi_{pp} =0.5\pi$,
where the copper-oxygen currents do flow in an ordered pattern, implying a tendency towards the $\Theta_{II}$ phase enhanced by the staggered flux terms.
However, the oxygen-oxygen currents cannot fully develop a $\Theta_{II}$ phase pattern, showing that the ground state still fails to sustain ordered circulating currents.
A variational cluster approximation study of a similar Hamiltonian reached the same conclusion.\cite{Lu_VCA_2012}
This can be attributed to the saddle point of the self-energy functional grand potential always being located at $\Delta \phi=0$; therefore, no spontaneous time-reversal symmetry-breaking phases are found in the variational calculations.

One should note that the current flow in the $\Theta_{II}$ phase assumes the shape of a skewed kagome lattice consisting of corner-sharing triangles [Fig. 7(a)].
On such a lattice, states with a $\mathbf{q}=0$ ordering vector could become the leading instability, where it would correspond to intra-unit cell ordering.  (For recent studies on the kagome Hubbard model see Refs. \onlinecite{Kiesel_PRB_2012,Kiesel_PRL_2013}, as well as related flux pattern approaches to stabilize spin liquids on the kagome geometry\cite{Ludwig_arXiv_2014}).
This implies that applying uni-axial stress or strain\cite{Chu_Science_2012, Hovnatan_PRL_2014} parallel to the anapole moment direction\cite{Matteo_PRB_2012,Varma_arXiv_2013} to distort the copper-oxygen plaquettes [Fig. 7(b)] could enhance the experimentally observed time-reversal symmetry-breaking signal, if it is indeed a result of the $\Theta_{II}$ phase.

This idea is further substantiated by ED calculations where the effect of lattice distortions is simulated by simultaneously increasing oxygen-oxygen hybridization along $(\bar{1}10)$ and decreasing it along (110).
The resulting correlation strengths are enhanced for currents flowing in the same directions as those in Fig. 7(a), whereas it is reduced for currents flowing in the ``wrong" directions.
Thus, in the limit that the (110) oxygen-oxygen hoppings are fully blocked and the hybridization pathways are exactly those on a kagome lattice, the staggered flux terms can stabilize the $\Theta_{II}$ phase, which still leaves open the ultimate question whether these currents could form spontaneously.

\section{Conclusion}

We have employed numerically exact techniques to study different proposals for the cuprate pseudogap.
These calculations rule out intra-unit-cell oxygen antiferromagnetism, as oxygen spin correlations are found to be too weak to explain the experimentally observed magnetic moment.
The calculations find no $\Theta_{II}$ phase stabilized in the ground state,
as the current-current correlations do not show any tendency towards such an order for the system sizes considered here.
Indeed, the $\Theta_{II}$ phase does not become the dominant order even when values for the charge-transfer gap $\Delta$, the on-site oxygen repulsion $U_{pp}$, or the inter-site interactions $V_{pd}$ and $V_{pp}$ are chosen to favor  circulating currents.
These results suggest that physics beyond the three-orbital model may be necessary to support a circulating current phase.
One variant of the orbital loop currents, proposed to explain the tilted magnetic moment seen by neutron scattering, is based on loop currents that circulate around the planar and apex oxygens.
To simulate this scenario, we have included additional apex oxygen $2p_z$ orbitals with a site energy $\epsilon_{p_z}$ that varies from 0 to 3.6 eV. However, none of the calculations exhibits the expected pattern of out-of-plane orbital loop currents.
We thereby conclude that the multi-orbital Hubbard model does not sustain an orbital loop current ground state for parameters relevant to cuprate superconductors.

Essentially, charge in the $\Theta_{II}$ phase would flow in a path resembling the structure of a skewed kagome lattice.
On such a lattice, circulating currents can be stabilized by including additional staggered flux terms.
The necessity of the staggered flux suggests that orbital loop currents do not occur naturally as a ground state in these models, but must be driven explicitly by terms with broken time-reversal symmetry.
This would also imply that applying uni-axial stress or strain parallel to the anapole moment could enhance the experimentally observed time-reversal symmetry-breaking signals, if they are indeed caused by spontaneous orbital loop currents. 

Given our findings, it may be necessary to turn to alternative explanations for the cuprate pseudogap, or to consider effects such as coexisting orders, impurity scattering, and extra orbital degrees of freedom.\cite{Yakovenko_arXiv_2014}
However, our study does not preclude the existence of orbital loop currents in other systems.\cite{Scagnoli_Science_2011,Matteo_PRB_2012, Varma_PRB_2013}  For these reasons, exploration of circulating current phases with spontaneously broken time-reversal symmetry will continue to be a topic of much interest.

The authors acknowledge helpful discussions with Chandra Varma, C\'{e}dric Weber, Arno Kampf, Douglas Scalapino, Sri Raghu, Steve Kivelson, Richard Scalettar, Chunjing Jia,  Alexander Kemper, Tsez\'{a}r Seman, and Jiun-Haw Chu.
Part of this research was supported by the U.S. Department of Energy (DOE), Office of Basic Energy Sciences, Division of Materials Sciences and Engineering, under Contract No. DE-AC02-76SF00515, SLAC National Accelerator Laboratory (SLAC), Stanford Institute for Materials and Energy Sciences.  Y.F.K. was supported by the Department of Defense (DOD) through the National Defense Science and Engineering Graduate Fellowship (NDSEG) Program and by the National Science Foundation (NSF) Graduate Research Fellowship under Grant No. 1147470.  C.C.C. is supported by the Aneesur Rahman Postdoctoral Fellowship at Argonne National Laboratory, operated under the U.S. DOE Contract No. DE-AC02-06CH11357.
The computational work was partially performed at the National Energy Research Scientific Computing Center (NERSC), supported by the U.S. DOE under Contract No. DE-AC02-05CH11231.

\bibliography{Pseudogap_Bib}

\begin{thebibliography}{89}
\expandafter\ifx\csname natexlab\endcsname\relax\def\natexlab#1{#1}\fi
\expandafter\ifx\csname bibnamefont\endcsname\relax
  \def\bibnamefont#1{#1}\fi
\expandafter\ifx\csname bibfnamefont\endcsname\relax
  \def\bibfnamefont#1{#1}\fi
\expandafter\ifx\csname citenamefont\endcsname\relax
  \def\citenamefont#1{#1}\fi
\expandafter\ifx\csname url\endcsname\relax
  \def\url#1{\texttt{#1}}\fi
\expandafter\ifx\csname urlprefix\endcsname\relax\def\urlprefix{URL }\fi
\providecommand{\bibinfo}[2]{#2}
\providecommand{\eprint}[2][]{\url{#2}}

\bibitem[{\citenamefont{Timusk and Statt}(1999)}]{Timusk_RPP_1999}
\bibinfo{author}{\bibfnamefont{T.}~\bibnamefont{Timusk}} \bibnamefont{and}
  \bibinfo{author}{\bibfnamefont{B.}~\bibnamefont{Statt}},
  \bibinfo{journal}{Rep. Prog. Phys.} \textbf{\bibinfo{volume}{62}},
  \bibinfo{pages}{61} (\bibinfo{year}{1999}).

\bibitem[{\citenamefont{Norman et~al.}(2005)\citenamefont{Norman, Pines, and
  Kallin}}]{Norman_2005}
\bibinfo{author}{\bibfnamefont{M.~R.} \bibnamefont{Norman}},
  \bibinfo{author}{\bibfnamefont{D.}~\bibnamefont{Pines}}, \bibnamefont{and}
  \bibinfo{author}{\bibfnamefont{C.}~\bibnamefont{Kallin}},
  \bibinfo{journal}{Adv. Phys.} \textbf{\bibinfo{volume}{54}},
  \bibinfo{pages}{715} (\bibinfo{year}{2005}).

\bibitem[{\citenamefont{Fradkin et~al.}(2014)\citenamefont{Fradkin, Kivelson,
  and Tranquada}}]{Fradkin_arXiv_2014}
\bibinfo{author}{\bibfnamefont{E.}~\bibnamefont{Fradkin}},
  \bibinfo{author}{\bibfnamefont{S.~A.} \bibnamefont{Kivelson}},
  \bibnamefont{and} \bibinfo{author}{\bibfnamefont{J.~M.}
  \bibnamefont{Tranquada}}, \bibinfo{journal}{arXiv:1407.4480}
  (\bibinfo{year}{2014}).

\bibitem[{\citenamefont{Hashimoto
  et~al.}(2014{\natexlab{a}})\citenamefont{Hashimoto, Vishik, He, Devereaux,
  and Shen}}]{Hashimoto_NatPhys_2014}
\bibinfo{author}{\bibfnamefont{M.}~\bibnamefont{Hashimoto}},
  \bibinfo{author}{\bibfnamefont{I.~M.} \bibnamefont{Vishik}},
  \bibinfo{author}{\bibfnamefont{R.-H.} \bibnamefont{He}},
  \bibinfo{author}{\bibfnamefont{T.~P.} \bibnamefont{Devereaux}},
  \bibnamefont{and} \bibinfo{author}{\bibfnamefont{Z.-X.} \bibnamefont{Shen}},
  \bibinfo{journal}{Nature Physics} \textbf{\bibinfo{volume}{10}},
  \bibinfo{pages}{483} (\bibinfo{year}{2014}{\natexlab{a}}).

\bibitem[{\citenamefont{Emery and Kivelson}(1995)}]{Emery_Nature_1995}
\bibinfo{author}{\bibfnamefont{V.}~\bibnamefont{Emery}} \bibnamefont{and}
  \bibinfo{author}{\bibfnamefont{S.}~\bibnamefont{Kivelson}},
  \bibinfo{journal}{Nature} \textbf{\bibinfo{volume}{374}},
  \bibinfo{pages}{434} (\bibinfo{year}{1995}).

\bibitem[{\citenamefont{Bourges and Sidis}(2011)}]{Bourges_Comptes_2011}
\bibinfo{author}{\bibfnamefont{P.}~\bibnamefont{Bourges}} \bibnamefont{and}
  \bibinfo{author}{\bibfnamefont{Y.}~\bibnamefont{Sidis}},
  \bibinfo{journal}{Comptes Rendus Physique} \textbf{\bibinfo{volume}{12}},
  \bibinfo{pages}{461} (\bibinfo{year}{2011}).

\bibitem[{\citenamefont{Varma}(1997)}]{Varma_PRB_1997}
\bibinfo{author}{\bibfnamefont{C.~M.} \bibnamefont{Varma}},
  \bibinfo{journal}{Phys. Rev. B} \textbf{\bibinfo{volume}{55}},
  \bibinfo{pages}{14554} (\bibinfo{year}{1997}).

\bibitem[{\citenamefont{Varma}(1999)}]{Varma_PRL_1999}
\bibinfo{author}{\bibfnamefont{C.~M.} \bibnamefont{Varma}},
  \bibinfo{journal}{Phys. Rev. Lett.} \textbf{\bibinfo{volume}{83}},
  \bibinfo{pages}{3538} (\bibinfo{year}{1999}).

\bibitem[{\citenamefont{Varma}(2006)}]{Varma_PRB_2006}
\bibinfo{author}{\bibfnamefont{C.~M.} \bibnamefont{Varma}},
  \bibinfo{journal}{Phys. Rev. B} \textbf{\bibinfo{volume}{73}},
  \bibinfo{pages}{155113} (\bibinfo{year}{2006}).

\bibitem[{\citenamefont{He and Varma}(2012)}]{Varma_PRB_2012}
\bibinfo{author}{\bibfnamefont{Y.}~\bibnamefont{He}} \bibnamefont{and}
  \bibinfo{author}{\bibfnamefont{C.~M.} \bibnamefont{Varma}},
  \bibinfo{journal}{Phys. Rev. B} \textbf{\bibinfo{volume}{86}},
  \bibinfo{pages}{035124} (\bibinfo{year}{2012}).

\bibitem[{\citenamefont{Weber et~al.}(2009)\citenamefont{Weber, L\"auchli,
  Mila, and Giamarchi}}]{Weber_PRL_2009}
\bibinfo{author}{\bibfnamefont{C.}~\bibnamefont{Weber}},
  \bibinfo{author}{\bibfnamefont{A.}~\bibnamefont{L\"auchli}},
  \bibinfo{author}{\bibfnamefont{F.}~\bibnamefont{Mila}}, \bibnamefont{and}
  \bibinfo{author}{\bibfnamefont{T.}~\bibnamefont{Giamarchi}},
  \bibinfo{journal}{Phys. Rev. Lett.} \textbf{\bibinfo{volume}{102}},
  \bibinfo{pages}{017005} (\bibinfo{year}{2009}).

\bibitem[{\citenamefont{Chakravarty et~al.}(2001)\citenamefont{Chakravarty,
  Laughlin, Morr, and Nayak}}]{Chakravarty_PRB_2001}
\bibinfo{author}{\bibfnamefont{S.}~\bibnamefont{Chakravarty}},
  \bibinfo{author}{\bibfnamefont{R.~B.} \bibnamefont{Laughlin}},
  \bibinfo{author}{\bibfnamefont{D.~K.} \bibnamefont{Morr}}, \bibnamefont{and}
  \bibinfo{author}{\bibfnamefont{C.}~\bibnamefont{Nayak}},
  \bibinfo{journal}{Phys. Rev. B} \textbf{\bibinfo{volume}{63}},
  \bibinfo{pages}{094503} (\bibinfo{year}{2001}).

\bibitem[{\citenamefont{Sau et~al.}(2013)\citenamefont{Sau, Mandal, Tewari, and
  Chakravarty}}]{Chakravarty_PRB_2013}
\bibinfo{author}{\bibfnamefont{J.~D.} \bibnamefont{Sau}},
  \bibinfo{author}{\bibfnamefont{I.}~\bibnamefont{Mandal}},
  \bibinfo{author}{\bibfnamefont{S.}~\bibnamefont{Tewari}}, \bibnamefont{and}
  \bibinfo{author}{\bibfnamefont{S.}~\bibnamefont{Chakravarty}},
  \bibinfo{journal}{Phys. Rev. B} \textbf{\bibinfo{volume}{87}},
  \bibinfo{pages}{224503} (\bibinfo{year}{2013}).

\bibitem[{\citenamefont{Kivelson et~al.}(2003)\citenamefont{Kivelson, Bindloss,
  Fradkin, Oganesyan, Tranquada, Kapitulnik, and Howald}}]{Kivelson_RMP_2003}
\bibinfo{author}{\bibfnamefont{S.~A.} \bibnamefont{Kivelson}},
  \bibinfo{author}{\bibfnamefont{I.~P.} \bibnamefont{Bindloss}},
  \bibinfo{author}{\bibfnamefont{E.}~\bibnamefont{Fradkin}},
  \bibinfo{author}{\bibfnamefont{V.}~\bibnamefont{Oganesyan}},
  \bibinfo{author}{\bibfnamefont{J.~M.} \bibnamefont{Tranquada}},
  \bibinfo{author}{\bibfnamefont{A.}~\bibnamefont{Kapitulnik}},
  \bibnamefont{and} \bibinfo{author}{\bibfnamefont{C.}~\bibnamefont{Howald}},
  \bibinfo{journal}{Rev. Mod. Phys.} \textbf{\bibinfo{volume}{75}},
  \bibinfo{pages}{1201} (\bibinfo{year}{2003}).

\bibitem[{\citenamefont{Lawler et~al.}(2010)\citenamefont{Lawler, Fujita, Lee,
  Schmidt, Kohsaka, Kim, Eisaki, Uchida, Davis, Sethna
  et~al.}}]{Lawler_Nature_2010}
\bibinfo{author}{\bibfnamefont{M.~J.} \bibnamefont{Lawler}},
  \bibinfo{author}{\bibfnamefont{K.}~\bibnamefont{Fujita}},
  \bibinfo{author}{\bibfnamefont{J.}~\bibnamefont{Lee}},
  \bibinfo{author}{\bibfnamefont{A.~R.} \bibnamefont{Schmidt}},
  \bibinfo{author}{\bibfnamefont{Y.}~\bibnamefont{Kohsaka}},
  \bibinfo{author}{\bibfnamefont{C.~K.} \bibnamefont{Kim}},
  \bibinfo{author}{\bibfnamefont{H.}~\bibnamefont{Eisaki}},
  \bibinfo{author}{\bibfnamefont{S.}~\bibnamefont{Uchida}},
  \bibinfo{author}{\bibfnamefont{J.~C.} \bibnamefont{Davis}},
  \bibinfo{author}{\bibfnamefont{J.~P.} \bibnamefont{Sethna}},
  \bibnamefont{et~al.}, \bibinfo{journal}{Nature}
  \textbf{\bibinfo{volume}{466}}, \bibinfo{pages}{347} (\bibinfo{year}{2010}).

\bibitem[{\citenamefont{Nie et~al.}(2014)\citenamefont{Nie, Tarjus, and
  Kivelson}}]{Nie_PNAS_2014}
\bibinfo{author}{\bibfnamefont{L.}~\bibnamefont{Nie}},
  \bibinfo{author}{\bibfnamefont{G.}~\bibnamefont{Tarjus}}, \bibnamefont{and}
  \bibinfo{author}{\bibfnamefont{S.~A.} \bibnamefont{Kivelson}},
  \bibinfo{journal}{Proceedings of the National Academy of Sciences}
  \textbf{\bibinfo{volume}{111}}, \bibinfo{pages}{7980} (\bibinfo{year}{2014}).

\bibitem[{\citenamefont{Chang et~al.}(2012)\citenamefont{Chang, Blackburn,
  Holmes, Christensen, Larsen, Mesot, Liang, Bonn, Hardy, Watenphul
  et~al.}}]{Chang_NatPhys_2012}
\bibinfo{author}{\bibfnamefont{J.}~\bibnamefont{Chang}},
  \bibinfo{author}{\bibfnamefont{E.}~\bibnamefont{Blackburn}},
  \bibinfo{author}{\bibfnamefont{A.~T.} \bibnamefont{Holmes}},
  \bibinfo{author}{\bibfnamefont{N.~B.} \bibnamefont{Christensen}},
  \bibinfo{author}{\bibfnamefont{J.}~\bibnamefont{Larsen}},
  \bibinfo{author}{\bibfnamefont{J.}~\bibnamefont{Mesot}},
  \bibinfo{author}{\bibfnamefont{R.}~\bibnamefont{Liang}},
  \bibinfo{author}{\bibfnamefont{D.~A.} \bibnamefont{Bonn}},
  \bibinfo{author}{\bibfnamefont{W.~N.} \bibnamefont{Hardy}},
  \bibinfo{author}{\bibfnamefont{A.}~\bibnamefont{Watenphul}},
  \bibnamefont{et~al.}, \bibinfo{journal}{Nature Phys.}
  \textbf{\bibinfo{volume}{8}}, \bibinfo{pages}{871} (\bibinfo{year}{2012}).

\bibitem[{\citenamefont{Ghiringhelli et~al.}(2012)\citenamefont{Ghiringhelli,
  Le~Tacon, Minola, Blanco-Canosa, Mazzoli, Brookes, De~Luca, Frano, Hawthorn,
  He et~al.}}]{Ghiringhelli_Science_2012}
\bibinfo{author}{\bibfnamefont{G.}~\bibnamefont{Ghiringhelli}},
  \bibinfo{author}{\bibfnamefont{M.}~\bibnamefont{Le~Tacon}},
  \bibinfo{author}{\bibfnamefont{M.}~\bibnamefont{Minola}},
  \bibinfo{author}{\bibfnamefont{S.}~\bibnamefont{Blanco-Canosa}},
  \bibinfo{author}{\bibfnamefont{C.}~\bibnamefont{Mazzoli}},
  \bibinfo{author}{\bibfnamefont{N.~B.} \bibnamefont{Brookes}},
  \bibinfo{author}{\bibfnamefont{G.~M.} \bibnamefont{De~Luca}},
  \bibinfo{author}{\bibfnamefont{A.}~\bibnamefont{Frano}},
  \bibinfo{author}{\bibfnamefont{D.~G.} \bibnamefont{Hawthorn}},
  \bibinfo{author}{\bibfnamefont{F.}~\bibnamefont{He}}, \bibnamefont{et~al.},
  \bibinfo{journal}{Science} \textbf{\bibinfo{volume}{337}},
  \bibinfo{pages}{821} (\bibinfo{year}{2012}).

\bibitem[{\citenamefont{Hashimoto et~al.}(2010)\citenamefont{Hashimoto, He,
  Tanaka, Testaud, Meevasana, Moore, Lu, Yao, Yoshida, Eisaki
  et~al.}}]{Hashimoto_NatPhys_2010}
\bibinfo{author}{\bibfnamefont{M.}~\bibnamefont{Hashimoto}},
  \bibinfo{author}{\bibfnamefont{R.-H.} \bibnamefont{He}},
  \bibinfo{author}{\bibfnamefont{K.}~\bibnamefont{Tanaka}},
  \bibinfo{author}{\bibfnamefont{J.-P.} \bibnamefont{Testaud}},
  \bibinfo{author}{\bibfnamefont{W.}~\bibnamefont{Meevasana}},
  \bibinfo{author}{\bibfnamefont{R.~G.} \bibnamefont{Moore}},
  \bibinfo{author}{\bibfnamefont{D.}~\bibnamefont{Lu}},
  \bibinfo{author}{\bibfnamefont{H.}~\bibnamefont{Yao}},
  \bibinfo{author}{\bibfnamefont{Y.}~\bibnamefont{Yoshida}},
  \bibinfo{author}{\bibfnamefont{H.}~\bibnamefont{Eisaki}},
  \bibnamefont{et~al.}, \bibinfo{journal}{Nature Physics}
  \textbf{\bibinfo{volume}{6}}, \bibinfo{pages}{414} (\bibinfo{year}{2010}).

\bibitem[{\citenamefont{Ma et~al.}(2008)\citenamefont{Ma, Pan, Niestemski,
  Neupane, Xu, Richard, Nakayama, Sato, Takahashi, Luo et~al.}}]{Ma_PRL_2008}
\bibinfo{author}{\bibfnamefont{J.-H.} \bibnamefont{Ma}},
  \bibinfo{author}{\bibfnamefont{Z.-H.} \bibnamefont{Pan}},
  \bibinfo{author}{\bibfnamefont{F.~C.} \bibnamefont{Niestemski}},
  \bibinfo{author}{\bibfnamefont{M.}~\bibnamefont{Neupane}},
  \bibinfo{author}{\bibfnamefont{Y.-M.} \bibnamefont{Xu}},
  \bibinfo{author}{\bibfnamefont{P.}~\bibnamefont{Richard}},
  \bibinfo{author}{\bibfnamefont{K.}~\bibnamefont{Nakayama}},
  \bibinfo{author}{\bibfnamefont{T.}~\bibnamefont{Sato}},
  \bibinfo{author}{\bibfnamefont{T.}~\bibnamefont{Takahashi}},
  \bibinfo{author}{\bibfnamefont{H.-Q.} \bibnamefont{Luo}},
  \bibnamefont{et~al.}, \bibinfo{journal}{Phys. Rev. Lett.}
  \textbf{\bibinfo{volume}{101}}, \bibinfo{pages}{207002}
  (\bibinfo{year}{2008}).

\bibitem[{\citenamefont{Vershinin et~al.}(2004)\citenamefont{Vershinin, Misra,
  Ono, Abe, Ando, and Yazdani}}]{Vershinin_Science_2004}
\bibinfo{author}{\bibfnamefont{M.}~\bibnamefont{Vershinin}},
  \bibinfo{author}{\bibfnamefont{S.}~\bibnamefont{Misra}},
  \bibinfo{author}{\bibfnamefont{S.}~\bibnamefont{Ono}},
  \bibinfo{author}{\bibfnamefont{Y.}~\bibnamefont{Abe}},
  \bibinfo{author}{\bibfnamefont{Y.}~\bibnamefont{Ando}}, \bibnamefont{and}
  \bibinfo{author}{\bibfnamefont{A.}~\bibnamefont{Yazdani}},
  \bibinfo{journal}{Science} \textbf{\bibinfo{volume}{303}},
  \bibinfo{pages}{1995} (\bibinfo{year}{2004}).

\bibitem[{\citenamefont{Wise et~al.}(2008)\citenamefont{Wise, Boyer,
  Chatterjee, Kondo, Takeuchi, Ikuta, Wang, and Hudson}}]{Wise_NatPhys_2008}
\bibinfo{author}{\bibfnamefont{W.}~\bibnamefont{Wise}},
  \bibinfo{author}{\bibfnamefont{M.}~\bibnamefont{Boyer}},
  \bibinfo{author}{\bibfnamefont{K.}~\bibnamefont{Chatterjee}},
  \bibinfo{author}{\bibfnamefont{T.}~\bibnamefont{Kondo}},
  \bibinfo{author}{\bibfnamefont{T.}~\bibnamefont{Takeuchi}},
  \bibinfo{author}{\bibfnamefont{H.}~\bibnamefont{Ikuta}},
  \bibinfo{author}{\bibfnamefont{Y.}~\bibnamefont{Wang}}, \bibnamefont{and}
  \bibinfo{author}{\bibfnamefont{E.}~\bibnamefont{Hudson}},
  \bibinfo{journal}{Nature Physics} \textbf{\bibinfo{volume}{4}},
  \bibinfo{pages}{696} (\bibinfo{year}{2008}).

\bibitem[{\citenamefont{Hashimoto
  et~al.}(2014{\natexlab{b}})\citenamefont{Hashimoto, Ghiringhelli, Lee,
  Dellea, Amorese, Mazzoli, Kummer, Brookes, Moritz, Yoshida
  et~al.}}]{Hashimoto_PRB_2014}
\bibinfo{author}{\bibfnamefont{M.}~\bibnamefont{Hashimoto}},
  \bibinfo{author}{\bibfnamefont{G.}~\bibnamefont{Ghiringhelli}},
  \bibinfo{author}{\bibfnamefont{W.-S.} \bibnamefont{Lee}},
  \bibinfo{author}{\bibfnamefont{G.}~\bibnamefont{Dellea}},
  \bibinfo{author}{\bibfnamefont{A.}~\bibnamefont{Amorese}},
  \bibinfo{author}{\bibfnamefont{C.}~\bibnamefont{Mazzoli}},
  \bibinfo{author}{\bibfnamefont{K.}~\bibnamefont{Kummer}},
  \bibinfo{author}{\bibfnamefont{N.~B.} \bibnamefont{Brookes}},
  \bibinfo{author}{\bibfnamefont{B.}~\bibnamefont{Moritz}},
  \bibinfo{author}{\bibfnamefont{Y.}~\bibnamefont{Yoshida}},
  \bibnamefont{et~al.}, \bibinfo{journal}{Phys. Rev. B}
  \textbf{\bibinfo{volume}{89}}, \bibinfo{pages}{220511}
  (\bibinfo{year}{2014}{\natexlab{b}}).

\bibitem[{\citenamefont{Vishik et~al.}(2012)\citenamefont{Vishik, Hashimoto,
  He, Lee, Schmitt, Lu, Moore, Zhang, Meevasana, Sasagawa
  et~al.}}]{Vishik_PNAS_2012}
\bibinfo{author}{\bibfnamefont{I.}~\bibnamefont{Vishik}},
  \bibinfo{author}{\bibfnamefont{M.}~\bibnamefont{Hashimoto}},
  \bibinfo{author}{\bibfnamefont{R.-H.} \bibnamefont{He}},
  \bibinfo{author}{\bibfnamefont{W.-S.} \bibnamefont{Lee}},
  \bibinfo{author}{\bibfnamefont{F.}~\bibnamefont{Schmitt}},
  \bibinfo{author}{\bibfnamefont{D.}~\bibnamefont{Lu}},
  \bibinfo{author}{\bibfnamefont{R.}~\bibnamefont{Moore}},
  \bibinfo{author}{\bibfnamefont{C.}~\bibnamefont{Zhang}},
  \bibinfo{author}{\bibfnamefont{W.}~\bibnamefont{Meevasana}},
  \bibinfo{author}{\bibfnamefont{T.}~\bibnamefont{Sasagawa}},
  \bibnamefont{et~al.}, \bibinfo{journal}{Proceedings of the National Academy
  of Sciences} \textbf{\bibinfo{volume}{109}}, \bibinfo{pages}{18332}
  (\bibinfo{year}{2012}).

\bibitem[{\citenamefont{Wu et~al.}(2011)\citenamefont{Wu, Mayaffre, Kr{\"a}mer,
  Horvati{\'c}, Berthier, Hardy, Liang, Bonn, and Julien}}]{Julien_Nature_2011}
\bibinfo{author}{\bibfnamefont{T.}~\bibnamefont{Wu}},
  \bibinfo{author}{\bibfnamefont{H.}~\bibnamefont{Mayaffre}},
  \bibinfo{author}{\bibfnamefont{S.}~\bibnamefont{Kr{\"a}mer}},
  \bibinfo{author}{\bibfnamefont{M.}~\bibnamefont{Horvati{\'c}}},
  \bibinfo{author}{\bibfnamefont{C.}~\bibnamefont{Berthier}},
  \bibinfo{author}{\bibfnamefont{W.}~\bibnamefont{Hardy}},
  \bibinfo{author}{\bibfnamefont{R.}~\bibnamefont{Liang}},
  \bibinfo{author}{\bibfnamefont{D.}~\bibnamefont{Bonn}}, \bibnamefont{and}
  \bibinfo{author}{\bibfnamefont{M.-H.} \bibnamefont{Julien}},
  \bibinfo{journal}{Nature} \textbf{\bibinfo{volume}{477}},
  \bibinfo{pages}{191} (\bibinfo{year}{2011}).

\bibitem[{\citenamefont{Wu et~al.}(2013)\citenamefont{Wu, Mayaffre, Kr{\"a}mer,
  Horvati{\'c}, Berthier, Kuhns, Reyes, Liang, Hardy, Bonn
  et~al.}}]{Julien_NatComm_2013}
\bibinfo{author}{\bibfnamefont{T.}~\bibnamefont{Wu}},
  \bibinfo{author}{\bibfnamefont{H.}~\bibnamefont{Mayaffre}},
  \bibinfo{author}{\bibfnamefont{S.}~\bibnamefont{Kr{\"a}mer}},
  \bibinfo{author}{\bibfnamefont{M.}~\bibnamefont{Horvati{\'c}}},
  \bibinfo{author}{\bibfnamefont{C.}~\bibnamefont{Berthier}},
  \bibinfo{author}{\bibfnamefont{P.~L.} \bibnamefont{Kuhns}},
  \bibinfo{author}{\bibfnamefont{A.~P.} \bibnamefont{Reyes}},
  \bibinfo{author}{\bibfnamefont{R.}~\bibnamefont{Liang}},
  \bibinfo{author}{\bibfnamefont{W.}~\bibnamefont{Hardy}},
  \bibinfo{author}{\bibfnamefont{D.}~\bibnamefont{Bonn}}, \bibnamefont{et~al.},
  \bibinfo{journal}{Nature Communications} \textbf{\bibinfo{volume}{4}}
  (\bibinfo{year}{2013}).

\bibitem[{\citenamefont{Wu et~al.}(2014)\citenamefont{Wu, Mayaffre, Kr{\"a}mer,
  Horvati{\'c}, Berthier, Hardy, Liang, Bonn, and Julien}}]{Julien_arXiv_2014}
\bibinfo{author}{\bibfnamefont{T.}~\bibnamefont{Wu}},
  \bibinfo{author}{\bibfnamefont{H.}~\bibnamefont{Mayaffre}},
  \bibinfo{author}{\bibfnamefont{S.}~\bibnamefont{Kr{\"a}mer}},
  \bibinfo{author}{\bibfnamefont{M.}~\bibnamefont{Horvati{\'c}}},
  \bibinfo{author}{\bibfnamefont{C.}~\bibnamefont{Berthier}},
  \bibinfo{author}{\bibfnamefont{W.}~\bibnamefont{Hardy}},
  \bibinfo{author}{\bibfnamefont{R.}~\bibnamefont{Liang}},
  \bibinfo{author}{\bibfnamefont{D.}~\bibnamefont{Bonn}}, \bibnamefont{and}
  \bibinfo{author}{\bibfnamefont{M.-H.} \bibnamefont{Julien}},
  \bibinfo{journal}{arXiv:1404.1617}  (\bibinfo{year}{2014}).

\bibitem[{\citenamefont{Xia et~al.}(2008)\citenamefont{Xia, Schemm, Deutscher,
  Kivelson, Bonn, Hardy, Liang, Siemons, Koster, Fejer et~al.}}]{Jia_PRL_2008}
\bibinfo{author}{\bibfnamefont{J.}~\bibnamefont{Xia}},
  \bibinfo{author}{\bibfnamefont{E.}~\bibnamefont{Schemm}},
  \bibinfo{author}{\bibfnamefont{G.}~\bibnamefont{Deutscher}},
  \bibinfo{author}{\bibfnamefont{S.~A.} \bibnamefont{Kivelson}},
  \bibinfo{author}{\bibfnamefont{D.~A.} \bibnamefont{Bonn}},
  \bibinfo{author}{\bibfnamefont{W.~N.} \bibnamefont{Hardy}},
  \bibinfo{author}{\bibfnamefont{R.}~\bibnamefont{Liang}},
  \bibinfo{author}{\bibfnamefont{W.}~\bibnamefont{Siemons}},
  \bibinfo{author}{\bibfnamefont{G.}~\bibnamefont{Koster}},
  \bibinfo{author}{\bibfnamefont{M.~M.} \bibnamefont{Fejer}},
  \bibnamefont{et~al.}, \bibinfo{journal}{Phys. Rev. Lett.}
  \textbf{\bibinfo{volume}{100}}, \bibinfo{pages}{127002}
  (\bibinfo{year}{2008}).

\bibitem[{\citenamefont{He et~al.}(2011)\citenamefont{He, Hashimoto,
  Karapetyan, Koralek, Hinton, Testaud, Nathan, Yoshida, Yao, Tanaka
  et~al.}}]{He_Science_2011}
\bibinfo{author}{\bibfnamefont{R.-H.} \bibnamefont{He}},
  \bibinfo{author}{\bibfnamefont{M.}~\bibnamefont{Hashimoto}},
  \bibinfo{author}{\bibfnamefont{H.}~\bibnamefont{Karapetyan}},
  \bibinfo{author}{\bibfnamefont{J.~D.} \bibnamefont{Koralek}},
  \bibinfo{author}{\bibfnamefont{J.~P.} \bibnamefont{Hinton}},
  \bibinfo{author}{\bibfnamefont{J.~P.} \bibnamefont{Testaud}},
  \bibinfo{author}{\bibfnamefont{V.}~\bibnamefont{Nathan}},
  \bibinfo{author}{\bibfnamefont{Y.}~\bibnamefont{Yoshida}},
  \bibinfo{author}{\bibfnamefont{H.}~\bibnamefont{Yao}},
  \bibinfo{author}{\bibfnamefont{K.}~\bibnamefont{Tanaka}},
  \bibnamefont{et~al.}, \bibinfo{journal}{Science}
  \textbf{\bibinfo{volume}{331}}, \bibinfo{pages}{1579} (\bibinfo{year}{2011}).

\bibitem[{\citenamefont{Karapetyan et~al.}(2012)\citenamefont{Karapetyan,
  H\"ucker, Gu, Tranquada, Fejer, Xia, and Kapitulnik}}]{Kapitulnik_PRL_2012}
\bibinfo{author}{\bibfnamefont{H.}~\bibnamefont{Karapetyan}},
  \bibinfo{author}{\bibfnamefont{M.}~\bibnamefont{H\"ucker}},
  \bibinfo{author}{\bibfnamefont{G.~D.} \bibnamefont{Gu}},
  \bibinfo{author}{\bibfnamefont{J.~M.} \bibnamefont{Tranquada}},
  \bibinfo{author}{\bibfnamefont{M.~M.} \bibnamefont{Fejer}},
  \bibinfo{author}{\bibfnamefont{J.}~\bibnamefont{Xia}}, \bibnamefont{and}
  \bibinfo{author}{\bibfnamefont{A.}~\bibnamefont{Kapitulnik}},
  \bibinfo{journal}{Phys. Rev. Lett.} \textbf{\bibinfo{volume}{109}},
  \bibinfo{pages}{147001} (\bibinfo{year}{2012}).

\bibitem[{\citenamefont{Hosur et~al.}(2013)\citenamefont{Hosur, Kapitulnik,
  Kivelson, Orenstein, and Raghu}}]{Hosur_PRB_2013}
\bibinfo{author}{\bibfnamefont{P.}~\bibnamefont{Hosur}},
  \bibinfo{author}{\bibfnamefont{A.}~\bibnamefont{Kapitulnik}},
  \bibinfo{author}{\bibfnamefont{S.~A.} \bibnamefont{Kivelson}},
  \bibinfo{author}{\bibfnamefont{J.}~\bibnamefont{Orenstein}},
  \bibnamefont{and} \bibinfo{author}{\bibfnamefont{S.}~\bibnamefont{Raghu}},
  \bibinfo{journal}{Phys. Rev. B} \textbf{\bibinfo{volume}{87}},
  \bibinfo{pages}{115116} (\bibinfo{year}{2013}).

\bibitem[{\citenamefont{Orenstein and Moore}(2013)}]{Orenstein_PRB_2013}
\bibinfo{author}{\bibfnamefont{J.}~\bibnamefont{Orenstein}} \bibnamefont{and}
  \bibinfo{author}{\bibfnamefont{J.~E.} \bibnamefont{Moore}},
  \bibinfo{journal}{Phys. Rev. B} \textbf{\bibinfo{volume}{87}},
  \bibinfo{pages}{165110} (\bibinfo{year}{2013}).

\bibitem[{\citenamefont{Sebastian et~al.}(2014)\citenamefont{Sebastian,
  Harrison, Balakirev, Alarawneh, Goddard, Liang, Bonn, Hardy, and
  Lonzarich}}]{Sebastian_Nature_2014}
\bibinfo{author}{\bibfnamefont{S.~E.} \bibnamefont{Sebastian}},
  \bibinfo{author}{\bibfnamefont{N.}~\bibnamefont{Harrison}},
  \bibinfo{author}{\bibfnamefont{F.~F.} \bibnamefont{Balakirev}},
  \bibinfo{author}{\bibfnamefont{M.~M.} \bibnamefont{Alarawneh}},
  \bibinfo{author}{\bibfnamefont{P.~A.} \bibnamefont{Goddard}},
  \bibinfo{author}{\bibfnamefont{R.}~\bibnamefont{Liang}},
  \bibinfo{author}{\bibfnamefont{D.~A.} \bibnamefont{Bonn}},
  \bibinfo{author}{\bibfnamefont{W.~N.} \bibnamefont{Hardy}}, \bibnamefont{and}
  \bibinfo{author}{\bibfnamefont{G.~G.} \bibnamefont{Lonzarich}},
  \bibinfo{journal}{Nature} \textbf{\bibinfo{volume}{511}}, \bibinfo{pages}{61}
  (\bibinfo{year}{2014}).

\bibitem[{\citenamefont{Karapetyan et~al.}(2014)\citenamefont{Karapetyan, Xia,
  H\"ucker, Gu, Tranquada, Fejer, and Kapitulnik}}]{Hovnatan_PRL_2014}
\bibinfo{author}{\bibfnamefont{H.}~\bibnamefont{Karapetyan}},
  \bibinfo{author}{\bibfnamefont{J.}~\bibnamefont{Xia}},
  \bibinfo{author}{\bibfnamefont{M.}~\bibnamefont{H\"ucker}},
  \bibinfo{author}{\bibfnamefont{G.~D.} \bibnamefont{Gu}},
  \bibinfo{author}{\bibfnamefont{J.~M.} \bibnamefont{Tranquada}},
  \bibinfo{author}{\bibfnamefont{M.~M.} \bibnamefont{Fejer}}, \bibnamefont{and}
  \bibinfo{author}{\bibfnamefont{A.}~\bibnamefont{Kapitulnik}},
  \bibinfo{journal}{Phys. Rev. Lett.} \textbf{\bibinfo{volume}{112}},
  \bibinfo{pages}{047003} (\bibinfo{year}{2014}),
  \urlprefix\url{http://link.aps.org/doi/10.1103/PhysRevLett.112.047003}.

\bibitem[{\citenamefont{Li et~al.}(2008)\citenamefont{Li, Bal{\'e}dent,
  Bari{\v{s}}i{\'c}, Cho, Fauqu{\'e}, Sidis, Yu, Zhao, Bourges, and
  Greven}}]{Li_Nature_2008}
\bibinfo{author}{\bibfnamefont{Y.}~\bibnamefont{Li}},
  \bibinfo{author}{\bibfnamefont{V.}~\bibnamefont{Bal{\'e}dent}},
  \bibinfo{author}{\bibfnamefont{N.}~\bibnamefont{Bari{\v{s}}i{\'c}}},
  \bibinfo{author}{\bibfnamefont{Y.}~\bibnamefont{Cho}},
  \bibinfo{author}{\bibfnamefont{B.}~\bibnamefont{Fauqu{\'e}}},
  \bibinfo{author}{\bibfnamefont{Y.}~\bibnamefont{Sidis}},
  \bibinfo{author}{\bibfnamefont{G.}~\bibnamefont{Yu}},
  \bibinfo{author}{\bibfnamefont{X.}~\bibnamefont{Zhao}},
  \bibinfo{author}{\bibfnamefont{P.}~\bibnamefont{Bourges}}, \bibnamefont{and}
  \bibinfo{author}{\bibfnamefont{M.}~\bibnamefont{Greven}},
  \bibinfo{journal}{Nature} \textbf{\bibinfo{volume}{455}},
  \bibinfo{pages}{372} (\bibinfo{year}{2008}).

\bibitem[{\citenamefont{Li et~al.}(2010)\citenamefont{Li, Bal{\'e}dent, Yu,
  Bari{\v{s}}i{\'c}, Hradil, Mole, Sidis, Steffens, Zhao, Bourges
  et~al.}}]{Li_Nature_2010}
\bibinfo{author}{\bibfnamefont{Y.}~\bibnamefont{Li}},
  \bibinfo{author}{\bibfnamefont{V.}~\bibnamefont{Bal{\'e}dent}},
  \bibinfo{author}{\bibfnamefont{G.}~\bibnamefont{Yu}},
  \bibinfo{author}{\bibfnamefont{N.}~\bibnamefont{Bari{\v{s}}i{\'c}}},
  \bibinfo{author}{\bibfnamefont{K.}~\bibnamefont{Hradil}},
  \bibinfo{author}{\bibfnamefont{R.~A.} \bibnamefont{Mole}},
  \bibinfo{author}{\bibfnamefont{Y.}~\bibnamefont{Sidis}},
  \bibinfo{author}{\bibfnamefont{P.}~\bibnamefont{Steffens}},
  \bibinfo{author}{\bibfnamefont{X.}~\bibnamefont{Zhao}},
  \bibinfo{author}{\bibfnamefont{P.}~\bibnamefont{Bourges}},
  \bibnamefont{et~al.}, \bibinfo{journal}{Nature}
  \textbf{\bibinfo{volume}{468}}, \bibinfo{pages}{283} (\bibinfo{year}{2010}).

\bibitem[{\citenamefont{Sidis and Bourges}(2013)}]{Sidis_arXiv_2013}
\bibinfo{author}{\bibfnamefont{Y.}~\bibnamefont{Sidis}} \bibnamefont{and}
  \bibinfo{author}{\bibfnamefont{P.}~\bibnamefont{Bourges}},
  \bibinfo{journal}{J. Phys.: Conf. Ser.} \textbf{\bibinfo{volume}{449}},
  \bibinfo{pages}{012012} (\bibinfo{year}{2013}).

\bibitem[{\citenamefont{Fauqu\'e et~al.}(2006)\citenamefont{Fauqu\'e, Sidis,
  Hinkov, Pailh\`es, Lin, Chaud, and Bourges}}]{Fauque_PRL_2006}
\bibinfo{author}{\bibfnamefont{B.}~\bibnamefont{Fauqu\'e}},
  \bibinfo{author}{\bibfnamefont{Y.}~\bibnamefont{Sidis}},
  \bibinfo{author}{\bibfnamefont{V.}~\bibnamefont{Hinkov}},
  \bibinfo{author}{\bibfnamefont{S.}~\bibnamefont{Pailh\`es}},
  \bibinfo{author}{\bibfnamefont{C.~T.} \bibnamefont{Lin}},
  \bibinfo{author}{\bibfnamefont{X.}~\bibnamefont{Chaud}}, \bibnamefont{and}
  \bibinfo{author}{\bibfnamefont{P.}~\bibnamefont{Bourges}},
  \bibinfo{journal}{Phys. Rev. Lett.} \textbf{\bibinfo{volume}{96}},
  \bibinfo{pages}{197001} (\bibinfo{year}{2006}).

\bibitem[{\citenamefont{Kaminski et~al.}(2002)\citenamefont{Kaminski,
  Rosenkranz, Fretwell, Campuzano, Li, Cullen, You, Olson, Varma, and
  H\"{o}chat}}]{Kaminski_Nature_2002}
\bibinfo{author}{\bibfnamefont{A.}~\bibnamefont{Kaminski}},
  \bibinfo{author}{\bibfnamefont{S.}~\bibnamefont{Rosenkranz}},
  \bibinfo{author}{\bibfnamefont{H.~M.} \bibnamefont{Fretwell}},
  \bibinfo{author}{\bibfnamefont{J.~C.} \bibnamefont{Campuzano}},
  \bibinfo{author}{\bibfnamefont{Z.}~\bibnamefont{Li}},
  \bibinfo{author}{\bibfnamefont{W.~G.} \bibnamefont{Cullen}},
  \bibinfo{author}{\bibfnamefont{H.}~\bibnamefont{You}},
  \bibinfo{author}{\bibfnamefont{C.~G.} \bibnamefont{Olson}},
  \bibinfo{author}{\bibfnamefont{C.~M.} \bibnamefont{Varma}}, \bibnamefont{and}
  \bibinfo{author}{\bibfnamefont{H.}~\bibnamefont{H\"{o}chat}},
  \bibinfo{journal}{Nature} \textbf{\bibinfo{volume}{416}},
  \bibinfo{pages}{610} (\bibinfo{year}{2002}).

\bibitem[{\citenamefont{Simon and Varma}(2002)}]{Simon_PRL_2002}
\bibinfo{author}{\bibfnamefont{M.~E.} \bibnamefont{Simon}} \bibnamefont{and}
  \bibinfo{author}{\bibfnamefont{C.~M.} \bibnamefont{Varma}},
  \bibinfo{journal}{Phys. Rev. Lett.} \textbf{\bibinfo{volume}{89}},
  \bibinfo{pages}{247003} (\bibinfo{year}{2002}).

\bibitem[{\citenamefont{Berg et~al.}(2008)\citenamefont{Berg, Chen, and
  Kivelson}}]{Berg_PRL_2008}
\bibinfo{author}{\bibfnamefont{E.}~\bibnamefont{Berg}},
  \bibinfo{author}{\bibfnamefont{C.-C.} \bibnamefont{Chen}}, \bibnamefont{and}
  \bibinfo{author}{\bibfnamefont{S.~A.} \bibnamefont{Kivelson}},
  \bibinfo{journal}{Phys. Rev. Lett.} \textbf{\bibinfo{volume}{100}},
  \bibinfo{pages}{027003} (\bibinfo{year}{2008}).

\bibitem[{\citenamefont{Allais and Senthil}(2012)}]{Allais_PRB_2012}
\bibinfo{author}{\bibfnamefont{A.}~\bibnamefont{Allais}} \bibnamefont{and}
  \bibinfo{author}{\bibfnamefont{T.}~\bibnamefont{Senthil}},
  \bibinfo{journal}{Phys. Rev. B} \textbf{\bibinfo{volume}{86}},
  \bibinfo{pages}{045118} (\bibinfo{year}{2012}).

\bibitem[{\citenamefont{Nielsen et~al.}(2012)\citenamefont{Nielsen, Atkinson,
  and Andersen}}]{Nielsen_PRB_2012}
\bibinfo{author}{\bibfnamefont{W.~H.~P.} \bibnamefont{Nielsen}},
  \bibinfo{author}{\bibfnamefont{W.~A.} \bibnamefont{Atkinson}},
  \bibnamefont{and} \bibinfo{author}{\bibfnamefont{B.~M.}
  \bibnamefont{Andersen}}, \bibinfo{journal}{Phys. Rev. B}
  \textbf{\bibinfo{volume}{86}}, \bibinfo{pages}{054510}
  (\bibinfo{year}{2012}).

\bibitem[{\citenamefont{Kivelson and Varma}(2012)}]{Kivelson_arXiv_2012}
\bibinfo{author}{\bibfnamefont{S.}~\bibnamefont{Kivelson}} \bibnamefont{and}
  \bibinfo{author}{\bibfnamefont{C.}~\bibnamefont{Varma}},
  \bibinfo{journal}{arXiv:1208.6498v1 [cond-mat.str-el]}
  (\bibinfo{year}{2012}).

\bibitem[{\citenamefont{Wang and Vafek}(2013)}]{Wang_PRB_2013}
\bibinfo{author}{\bibfnamefont{L.}~\bibnamefont{Wang}} \bibnamefont{and}
  \bibinfo{author}{\bibfnamefont{O.}~\bibnamefont{Vafek}},
  \bibinfo{journal}{Phys. Rev. B} \textbf{\bibinfo{volume}{88}},
  \bibinfo{pages}{024506} (\bibinfo{year}{2013}).

\bibitem[{\citenamefont{Aji et~al.}(2013)\citenamefont{Aji, He, and
  Varma}}]{Aji_PRB_2013}
\bibinfo{author}{\bibfnamefont{V.}~\bibnamefont{Aji}},
  \bibinfo{author}{\bibfnamefont{Y.}~\bibnamefont{He}}, \bibnamefont{and}
  \bibinfo{author}{\bibfnamefont{C.~M.} \bibnamefont{Varma}},
  \bibinfo{journal}{Phys. Rev. B} \textbf{\bibinfo{volume}{87}},
  \bibinfo{pages}{174518} (\bibinfo{year}{2013}).

\bibitem[{\citenamefont{Str\"assle et~al.}(2011)\citenamefont{Str\"assle,
  Graneli, Mali, Roos, and Keller}}]{Strassle_PRL_2011}
\bibinfo{author}{\bibfnamefont{S.}~\bibnamefont{Str\"assle}},
  \bibinfo{author}{\bibfnamefont{B.}~\bibnamefont{Graneli}},
  \bibinfo{author}{\bibfnamefont{M.}~\bibnamefont{Mali}},
  \bibinfo{author}{\bibfnamefont{J.}~\bibnamefont{Roos}}, \bibnamefont{and}
  \bibinfo{author}{\bibfnamefont{H.}~\bibnamefont{Keller}},
  \bibinfo{journal}{Phys. Rev. Lett.} \textbf{\bibinfo{volume}{106}},
  \bibinfo{pages}{097003} (\bibinfo{year}{2011}).

\bibitem[{\citenamefont{Lederer and Kivelson}(2012)}]{Lederer_PRB_2012}
\bibinfo{author}{\bibfnamefont{S.}~\bibnamefont{Lederer}} \bibnamefont{and}
  \bibinfo{author}{\bibfnamefont{S.~A.} \bibnamefont{Kivelson}},
  \bibinfo{journal}{Phys. Rev. B} \textbf{\bibinfo{volume}{85}},
  \bibinfo{pages}{155130} (\bibinfo{year}{2012}).

\bibitem[{\citenamefont{Mounce et~al.}(2013)\citenamefont{Mounce, Oh, Lee,
  Halperin, Reyes, Kuhns, Chan, Dorow, Ji, Xia et~al.}}]{Mounce_PRL_2013}
\bibinfo{author}{\bibfnamefont{A.~M.} \bibnamefont{Mounce}},
  \bibinfo{author}{\bibfnamefont{S.}~\bibnamefont{Oh}},
  \bibinfo{author}{\bibfnamefont{J.~A.} \bibnamefont{Lee}},
  \bibinfo{author}{\bibfnamefont{W.~P.} \bibnamefont{Halperin}},
  \bibinfo{author}{\bibfnamefont{A.~P.} \bibnamefont{Reyes}},
  \bibinfo{author}{\bibfnamefont{P.~L.} \bibnamefont{Kuhns}},
  \bibinfo{author}{\bibfnamefont{M.~K.} \bibnamefont{Chan}},
  \bibinfo{author}{\bibfnamefont{C.}~\bibnamefont{Dorow}},
  \bibinfo{author}{\bibfnamefont{L.}~\bibnamefont{Ji}},
  \bibinfo{author}{\bibfnamefont{D.}~\bibnamefont{Xia}}, \bibnamefont{et~al.},
  \bibinfo{journal}{Phys. Rev. Lett.} \textbf{\bibinfo{volume}{111}},
  \bibinfo{pages}{187003} (\bibinfo{year}{2013}).

\bibitem[{\citenamefont{Sonier et~al.}(2009)\citenamefont{Sonier, Pacradouni,
  Sabok-Sayr, Hardy, Bonn, Liang, and Mook}}]{Sonier_PRL_2009}
\bibinfo{author}{\bibfnamefont{J.~E.} \bibnamefont{Sonier}},
  \bibinfo{author}{\bibfnamefont{V.}~\bibnamefont{Pacradouni}},
  \bibinfo{author}{\bibfnamefont{S.~A.} \bibnamefont{Sabok-Sayr}},
  \bibinfo{author}{\bibfnamefont{W.~N.} \bibnamefont{Hardy}},
  \bibinfo{author}{\bibfnamefont{D.~A.} \bibnamefont{Bonn}},
  \bibinfo{author}{\bibfnamefont{R.}~\bibnamefont{Liang}}, \bibnamefont{and}
  \bibinfo{author}{\bibfnamefont{H.~A.} \bibnamefont{Mook}},
  \bibinfo{journal}{Phys. Rev. Lett.} \textbf{\bibinfo{volume}{103}},
  \bibinfo{pages}{167002} (\bibinfo{year}{2009}).

\bibitem[{\citenamefont{Huang et~al.}(2012)\citenamefont{Huang, Pacradouni,
  Kennett, Komiya, and Sonier}}]{Huang_PRB_2012}
\bibinfo{author}{\bibfnamefont{W.}~\bibnamefont{Huang}},
  \bibinfo{author}{\bibfnamefont{V.}~\bibnamefont{Pacradouni}},
  \bibinfo{author}{\bibfnamefont{M.~P.} \bibnamefont{Kennett}},
  \bibinfo{author}{\bibfnamefont{S.}~\bibnamefont{Komiya}}, \bibnamefont{and}
  \bibinfo{author}{\bibfnamefont{J.~E.} \bibnamefont{Sonier}},
  \bibinfo{journal}{Phys. Rev. B} \textbf{\bibinfo{volume}{85}},
  \bibinfo{pages}{104527} (\bibinfo{year}{2012}).

\bibitem[{\citenamefont{Storchak et~al.}(2014)\citenamefont{Storchak, Brewer,
  Eshchenko, Mengyan, Parfenov, Tokmachev, Dosanjh, and
  Barilo}}]{Storchak_arXiv_2014}
\bibinfo{author}{\bibfnamefont{V.~G.} \bibnamefont{Storchak}},
  \bibinfo{author}{\bibfnamefont{J.~H.} \bibnamefont{Brewer}},
  \bibinfo{author}{\bibfnamefont{D.~G.} \bibnamefont{Eshchenko}},
  \bibinfo{author}{\bibfnamefont{P.~W.} \bibnamefont{Mengyan}},
  \bibinfo{author}{\bibfnamefont{O.~E.} \bibnamefont{Parfenov}},
  \bibinfo{author}{\bibfnamefont{A.~M.} \bibnamefont{Tokmachev}},
  \bibinfo{author}{\bibfnamefont{P.}~\bibnamefont{Dosanjh}}, \bibnamefont{and}
  \bibinfo{author}{\bibfnamefont{S.~N.} \bibnamefont{Barilo}},
  \bibinfo{journal}{arXiv:1407.3570}  (\bibinfo{year}{2014}).

\bibitem[{\citenamefont{Weber et~al.}(2014)\citenamefont{Weber, Giamarchi, and
  Varma}}]{Weber_PRL_2014}
\bibinfo{author}{\bibfnamefont{C.}~\bibnamefont{Weber}},
  \bibinfo{author}{\bibfnamefont{T.}~\bibnamefont{Giamarchi}},
  \bibnamefont{and} \bibinfo{author}{\bibfnamefont{C.~M.} \bibnamefont{Varma}},
  \bibinfo{journal}{Phys. Rev. Lett.} \textbf{\bibinfo{volume}{112}},
  \bibinfo{pages}{117001} (\bibinfo{year}{2014}).

\bibitem[{\citenamefont{Chudzinski et~al.}(2008)\citenamefont{Chudzinski,
  Gabay, and Giamarchi}}]{Giamarchi_Bosonization_2008}
\bibinfo{author}{\bibfnamefont{P.}~\bibnamefont{Chudzinski}},
  \bibinfo{author}{\bibfnamefont{M.}~\bibnamefont{Gabay}}, \bibnamefont{and}
  \bibinfo{author}{\bibfnamefont{T.}~\bibnamefont{Giamarchi}},
  \bibinfo{journal}{Phys. Rev. B} \textbf{\bibinfo{volume}{78}},
  \bibinfo{pages}{075124} (\bibinfo{year}{2008}).

\bibitem[{\citenamefont{Fischer and Kim}(2011)}]{Fischer_MFT_2011}
\bibinfo{author}{\bibfnamefont{M.~H.} \bibnamefont{Fischer}} \bibnamefont{and}
  \bibinfo{author}{\bibfnamefont{E.-A.} \bibnamefont{Kim}},
  \bibinfo{journal}{Phys. Rev. B} \textbf{\bibinfo{volume}{84}},
  \bibinfo{pages}{144502} (\bibinfo{year}{2011}).

\bibitem[{\citenamefont{Greiter and Thomale}(2007)}]{Thomale_PRL_2007}
\bibinfo{author}{\bibfnamefont{M.}~\bibnamefont{Greiter}} \bibnamefont{and}
  \bibinfo{author}{\bibfnamefont{R.}~\bibnamefont{Thomale}},
  \bibinfo{journal}{Phys. Rev. Lett.} \textbf{\bibinfo{volume}{99}},
  \bibinfo{pages}{027005} (\bibinfo{year}{2007}).

\bibitem[{\citenamefont{Thomale and Greiter}(2008)}]{Thomale_PRB_2008}
\bibinfo{author}{\bibfnamefont{R.}~\bibnamefont{Thomale}} \bibnamefont{and}
  \bibinfo{author}{\bibfnamefont{M.}~\bibnamefont{Greiter}},
  \bibinfo{journal}{Phys. Rev. B} \textbf{\bibinfo{volume}{77}},
  \bibinfo{pages}{094511} (\bibinfo{year}{2008}).

\bibitem[{\citenamefont{Lu et~al.}(2012)\citenamefont{Lu, Chioncel, and
  Arrigoni}}]{Lu_VCA_2012}
\bibinfo{author}{\bibfnamefont{X.}~\bibnamefont{Lu}},
  \bibinfo{author}{\bibfnamefont{L.}~\bibnamefont{Chioncel}}, \bibnamefont{and}
  \bibinfo{author}{\bibfnamefont{E.}~\bibnamefont{Arrigoni}},
  \bibinfo{journal}{Physical Review B} \textbf{\bibinfo{volume}{85}},
  \bibinfo{pages}{125117} (\bibinfo{year}{2012}).

\bibitem[{\citenamefont{Nishimoto et~al.}(2009)\citenamefont{Nishimoto,
  Jeckelmann, and Scalapino}}]{Scalapino_DMRG_2009}
\bibinfo{author}{\bibfnamefont{S.}~\bibnamefont{Nishimoto}},
  \bibinfo{author}{\bibfnamefont{E.}~\bibnamefont{Jeckelmann}},
  \bibnamefont{and} \bibinfo{author}{\bibfnamefont{D.~J.}
  \bibnamefont{Scalapino}}, \bibinfo{journal}{Phys. Rev. B}
  \textbf{\bibinfo{volume}{79}}, \bibinfo{pages}{205115}
  (\bibinfo{year}{2009}).

\bibitem[{\citenamefont{Hybertsen et~al.}(1989)\citenamefont{Hybertsen,
  Schl\"uter, and Christensen}}]{Hybertsen_PRB_1989}
\bibinfo{author}{\bibfnamefont{M.~S.} \bibnamefont{Hybertsen}},
  \bibinfo{author}{\bibfnamefont{M.}~\bibnamefont{Schl\"uter}},
  \bibnamefont{and} \bibinfo{author}{\bibfnamefont{N.~E.}
  \bibnamefont{Christensen}}, \bibinfo{journal}{Phys. Rev. B}
  \textbf{\bibinfo{volume}{39}}, \bibinfo{pages}{9028} (\bibinfo{year}{1989}).

\bibitem[{\citenamefont{Chu et~al.}(2012)\citenamefont{Chu, Kuo, Analytis, and
  Fisher}}]{Chu_Science_2012}
\bibinfo{author}{\bibfnamefont{J.-H.} \bibnamefont{Chu}},
  \bibinfo{author}{\bibfnamefont{H.-H.} \bibnamefont{Kuo}},
  \bibinfo{author}{\bibfnamefont{J.~G.} \bibnamefont{Analytis}},
  \bibnamefont{and} \bibinfo{author}{\bibfnamefont{I.~R.}
  \bibnamefont{Fisher}}, \bibinfo{journal}{Science}
  \textbf{\bibinfo{volume}{337}}, \bibinfo{pages}{710} (\bibinfo{year}{2012}).

\bibitem[{\citenamefont{Dagotto}(1994)}]{Dagotto_RMP_1994}
\bibinfo{author}{\bibfnamefont{E.}~\bibnamefont{Dagotto}},
  \bibinfo{journal}{Rev. Mod. Phys.} \textbf{\bibinfo{volume}{66}},
  \bibinfo{pages}{763} (\bibinfo{year}{1994}).

\bibitem[{\citenamefont{Mattheiss}(1987)}]{Mattheiss_PRL_1987}
\bibinfo{author}{\bibfnamefont{L.~F.} \bibnamefont{Mattheiss}},
  \bibinfo{journal}{Phys. Rev. Lett.} \textbf{\bibinfo{volume}{58}},
  \bibinfo{pages}{1028} (\bibinfo{year}{1987}).

\bibitem[{\citenamefont{Varma et~al.}(1987)\citenamefont{Varma, Schmitt-Rink,
  and Abrahams}}]{Varma_SSC_1987}
\bibinfo{author}{\bibfnamefont{C.~M.} \bibnamefont{Varma}},
  \bibinfo{author}{\bibfnamefont{S.}~\bibnamefont{Schmitt-Rink}},
  \bibnamefont{and} \bibinfo{author}{\bibfnamefont{E.}~\bibnamefont{Abrahams}},
  \bibinfo{journal}{Solid State Commun.} \textbf{\bibinfo{volume}{62}},
  \bibinfo{pages}{681} (\bibinfo{year}{1987}).

\bibitem[{\citenamefont{Emery}(1987)}]{Emery_PRL_1987}
\bibinfo{author}{\bibfnamefont{V.~J.} \bibnamefont{Emery}},
  \bibinfo{journal}{Phys. Rev. Lett.} \textbf{\bibinfo{volume}{58}},
  \bibinfo{pages}{2794} (\bibinfo{year}{1987}).

\bibitem[{\citenamefont{Ohta et~al.}(1991)\citenamefont{Ohta, Tohyama, and
  Maekawa}}]{Ohta_PRB_1991}
\bibinfo{author}{\bibfnamefont{Y.}~\bibnamefont{Ohta}},
  \bibinfo{author}{\bibfnamefont{T.}~\bibnamefont{Tohyama}}, \bibnamefont{and}
  \bibinfo{author}{\bibfnamefont{S.}~\bibnamefont{Maekawa}},
  \bibinfo{journal}{Phys. Rev. B} \textbf{\bibinfo{volume}{43}},
  \bibinfo{pages}{2968} (\bibinfo{year}{1991}).

\bibitem[{\citenamefont{Johnston et~al.}(2009)\citenamefont{Johnston, Vernay,
  and Devereaux}}]{Johnston_EPL_2009}
\bibinfo{author}{\bibfnamefont{S.}~\bibnamefont{Johnston}},
  \bibinfo{author}{\bibfnamefont{F.}~\bibnamefont{Vernay}}, \bibnamefont{and}
  \bibinfo{author}{\bibfnamefont{T.}~\bibnamefont{Devereaux}},
  \bibinfo{journal}{EPL (Europhysics Letters)} \textbf{\bibinfo{volume}{86}},
  \bibinfo{pages}{37007} (\bibinfo{year}{2009}).

\bibitem[{\citenamefont{Czy\ifmmode~\dot{z}\else \.{z}\fi{}yk and
  Sawatzky}(1994)}]{Czyzyk_PRB_1994}
\bibinfo{author}{\bibfnamefont{M.~T.} \bibnamefont{Czy\ifmmode~\dot{z}\else
  \.{z}\fi{}yk}} \bibnamefont{and} \bibinfo{author}{\bibfnamefont{G.~A.}
  \bibnamefont{Sawatzky}}, \bibinfo{journal}{Phys. Rev. B}
  \textbf{\bibinfo{volume}{49}}, \bibinfo{pages}{14211} (\bibinfo{year}{1994}).

\bibitem[{\citenamefont{Lehoucq et~al.}(1998)\citenamefont{Lehoucq, Sorensen,
  and Yang}}]{ARPACK}
\bibinfo{author}{\bibfnamefont{R.~B.} \bibnamefont{Lehoucq}},
  \bibinfo{author}{\bibfnamefont{D.~C.} \bibnamefont{Sorensen}},
  \bibnamefont{and} \bibinfo{author}{\bibfnamefont{C.}~\bibnamefont{Yang}},
  \emph{\bibinfo{title}{ARPACK Users' Guide: Solution of Large-Scale Eigenvalue
  Problems with Implicitly Restarted Arnoldi Methods}}
  (\bibinfo{publisher}{SIAM}, \bibinfo{address}{Philadelphia},
  \bibinfo{year}{1998}).

\bibitem[{\citenamefont{Blankenbecler et~al.}(1981)\citenamefont{Blankenbecler,
  Scalapino, and Sugar}}]{BSS_PRD_1981}
\bibinfo{author}{\bibfnamefont{R.}~\bibnamefont{Blankenbecler}},
  \bibinfo{author}{\bibfnamefont{D.~J.} \bibnamefont{Scalapino}},
  \bibnamefont{and} \bibinfo{author}{\bibfnamefont{R.~L.} \bibnamefont{Sugar}},
  \bibinfo{journal}{Phys. Rev. D} \textbf{\bibinfo{volume}{24}},
  \bibinfo{pages}{2278} (\bibinfo{year}{1981}).

\bibitem[{\citenamefont{Dopf et~al.}(1990)\citenamefont{Dopf, Muramatsu, and
  Hanke}}]{Dopf_PRB_1990}
\bibinfo{author}{\bibfnamefont{G.}~\bibnamefont{Dopf}},
  \bibinfo{author}{\bibfnamefont{A.}~\bibnamefont{Muramatsu}},
  \bibnamefont{and} \bibinfo{author}{\bibfnamefont{W.}~\bibnamefont{Hanke}},
  \bibinfo{journal}{Phys. Rev. B} \textbf{\bibinfo{volume}{41}},
  \bibinfo{pages}{9264} (\bibinfo{year}{1990}).

\bibitem[{\citenamefont{Scalettar et~al.}(1991)\citenamefont{Scalettar,
  Scalapino, Sugar, and White}}]{Scalettar_PRB_1991}
\bibinfo{author}{\bibfnamefont{R.~T.} \bibnamefont{Scalettar}},
  \bibinfo{author}{\bibfnamefont{D.~J.} \bibnamefont{Scalapino}},
  \bibinfo{author}{\bibfnamefont{R.~L.} \bibnamefont{Sugar}}, \bibnamefont{and}
  \bibinfo{author}{\bibfnamefont{S.~R.} \bibnamefont{White}},
  \bibinfo{journal}{Phys. Rev. B} \textbf{\bibinfo{volume}{44}},
  \bibinfo{pages}{770} (\bibinfo{year}{1991}).

\bibitem[{\citenamefont{Dopf et~al.}(1992)\citenamefont{Dopf, Muramatsu, and
  Hanke}}]{Dopf_PRL_1992}
\bibinfo{author}{\bibfnamefont{G.}~\bibnamefont{Dopf}},
  \bibinfo{author}{\bibfnamefont{A.}~\bibnamefont{Muramatsu}},
  \bibnamefont{and} \bibinfo{author}{\bibfnamefont{W.}~\bibnamefont{Hanke}},
  \bibinfo{journal}{Phys. Rev. Lett.} \textbf{\bibinfo{volume}{68}},
  \bibinfo{pages}{353} (\bibinfo{year}{1992}).

\bibitem[{\citenamefont{Moritz et~al.}(2009)\citenamefont{Moritz, Schmitt,
  Meevasana, Johnston, Motoyama, Greven, Lu, Kim, Scalettar, Shen
  et~al.}}]{Moritz_NJOP_2009}
\bibinfo{author}{\bibfnamefont{B.}~\bibnamefont{Moritz}},
  \bibinfo{author}{\bibfnamefont{F.}~\bibnamefont{Schmitt}},
  \bibinfo{author}{\bibfnamefont{W.}~\bibnamefont{Meevasana}},
  \bibinfo{author}{\bibfnamefont{S.}~\bibnamefont{Johnston}},
  \bibinfo{author}{\bibfnamefont{E.}~\bibnamefont{Motoyama}},
  \bibinfo{author}{\bibfnamefont{M.}~\bibnamefont{Greven}},
  \bibinfo{author}{\bibfnamefont{D.}~\bibnamefont{Lu}},
  \bibinfo{author}{\bibfnamefont{C.}~\bibnamefont{Kim}},
  \bibinfo{author}{\bibfnamefont{R.}~\bibnamefont{Scalettar}},
  \bibinfo{author}{\bibfnamefont{Z.}~\bibnamefont{Shen}}, \bibnamefont{et~al.},
  \bibinfo{journal}{New Journal of Physics} \textbf{\bibinfo{volume}{11}},
  \bibinfo{pages}{093020} (\bibinfo{year}{2009}).

\bibitem[{\citenamefont{Chen et~al.}(2010)\citenamefont{Chen, Moritz, Vernay,
  Hancock, Johnston, Jia, Chabot-Couture, Greven, Elfimov, Sawatzky
  et~al.}}]{Chen_PRL_2010}
\bibinfo{author}{\bibfnamefont{C.-C.} \bibnamefont{Chen}},
  \bibinfo{author}{\bibfnamefont{B.}~\bibnamefont{Moritz}},
  \bibinfo{author}{\bibfnamefont{F.}~\bibnamefont{Vernay}},
  \bibinfo{author}{\bibfnamefont{J.~N.} \bibnamefont{Hancock}},
  \bibinfo{author}{\bibfnamefont{S.}~\bibnamefont{Johnston}},
  \bibinfo{author}{\bibfnamefont{C.~J.} \bibnamefont{Jia}},
  \bibinfo{author}{\bibfnamefont{G.}~\bibnamefont{Chabot-Couture}},
  \bibinfo{author}{\bibfnamefont{M.}~\bibnamefont{Greven}},
  \bibinfo{author}{\bibfnamefont{I.}~\bibnamefont{Elfimov}},
  \bibinfo{author}{\bibfnamefont{G.~A.} \bibnamefont{Sawatzky}},
  \bibnamefont{et~al.}, \bibinfo{journal}{Phys. Rev. Lett.}
  \textbf{\bibinfo{volume}{105}}, \bibinfo{pages}{177401}
  (\bibinfo{year}{2010}).

\bibitem[{\citenamefont{Moritz et~al.}(2011)\citenamefont{Moritz, Johnston,
  Devereaux, Muschler, Prestel, Hackl, Lambacher, Erb, Komiya, and
  Ando}}]{Moritz_PRB_2011}
\bibinfo{author}{\bibfnamefont{B.}~\bibnamefont{Moritz}},
  \bibinfo{author}{\bibfnamefont{S.}~\bibnamefont{Johnston}},
  \bibinfo{author}{\bibfnamefont{T.~P.} \bibnamefont{Devereaux}},
  \bibinfo{author}{\bibfnamefont{B.}~\bibnamefont{Muschler}},
  \bibinfo{author}{\bibfnamefont{W.}~\bibnamefont{Prestel}},
  \bibinfo{author}{\bibfnamefont{R.}~\bibnamefont{Hackl}},
  \bibinfo{author}{\bibfnamefont{M.}~\bibnamefont{Lambacher}},
  \bibinfo{author}{\bibfnamefont{A.}~\bibnamefont{Erb}},
  \bibinfo{author}{\bibfnamefont{S.}~\bibnamefont{Komiya}}, \bibnamefont{and}
  \bibinfo{author}{\bibfnamefont{Y.}~\bibnamefont{Ando}},
  \bibinfo{journal}{Phys. Rev. B} \textbf{\bibinfo{volume}{84}},
  \bibinfo{pages}{235114} (\bibinfo{year}{2011}).

\bibitem[{\citenamefont{Jia et~al.}(2012)\citenamefont{Jia, Chen, Sorini,
  Moritz, and Devereaux}}]{Jia_NJOP_2012}
\bibinfo{author}{\bibfnamefont{C.}~\bibnamefont{Jia}},
  \bibinfo{author}{\bibfnamefont{C.}~\bibnamefont{Chen}},
  \bibinfo{author}{\bibfnamefont{A.}~\bibnamefont{Sorini}},
  \bibinfo{author}{\bibfnamefont{B.}~\bibnamefont{Moritz}}, \bibnamefont{and}
  \bibinfo{author}{\bibfnamefont{T.}~\bibnamefont{Devereaux}},
  \bibinfo{journal}{New Journal of Physics} \textbf{\bibinfo{volume}{14}},
  \bibinfo{pages}{113038} (\bibinfo{year}{2012}).

\bibitem[{\citenamefont{Chen et~al.}(2013)\citenamefont{Chen, Sentef, Kung,
  Jia, Thomale, Moritz, Kampf, and Devereaux}}]{Chen_PRB_2013}
\bibinfo{author}{\bibfnamefont{C.-C.} \bibnamefont{Chen}},
  \bibinfo{author}{\bibfnamefont{M.}~\bibnamefont{Sentef}},
  \bibinfo{author}{\bibfnamefont{Y.~F.} \bibnamefont{Kung}},
  \bibinfo{author}{\bibfnamefont{C.~J.} \bibnamefont{Jia}},
  \bibinfo{author}{\bibfnamefont{R.}~\bibnamefont{Thomale}},
  \bibinfo{author}{\bibfnamefont{B.}~\bibnamefont{Moritz}},
  \bibinfo{author}{\bibfnamefont{A.~P.} \bibnamefont{Kampf}}, \bibnamefont{and}
  \bibinfo{author}{\bibfnamefont{T.~P.} \bibnamefont{Devereaux}},
  \bibinfo{journal}{Phys. Rev. B} \textbf{\bibinfo{volume}{87}},
  \bibinfo{pages}{165144} (\bibinfo{year}{2013}).

\bibitem[{\citenamefont{Jia et~al.}(2014)\citenamefont{Jia, Nowadnick,
  Wohlfeld, Kung, Chen, Johnston, Tohyama, Moritz, and
  Devereaux}}]{Jia_NatComm_2014}
\bibinfo{author}{\bibfnamefont{C.}~\bibnamefont{Jia}},
  \bibinfo{author}{\bibfnamefont{E.}~\bibnamefont{Nowadnick}},
  \bibinfo{author}{\bibfnamefont{K.}~\bibnamefont{Wohlfeld}},
  \bibinfo{author}{\bibfnamefont{Y.}~\bibnamefont{Kung}},
  \bibinfo{author}{\bibfnamefont{C.-C.} \bibnamefont{Chen}},
  \bibinfo{author}{\bibfnamefont{S.}~\bibnamefont{Johnston}},
  \bibinfo{author}{\bibfnamefont{T.}~\bibnamefont{Tohyama}},
  \bibinfo{author}{\bibfnamefont{B.}~\bibnamefont{Moritz}}, \bibnamefont{and}
  \bibinfo{author}{\bibfnamefont{T.}~\bibnamefont{Devereaux}},
  \bibinfo{journal}{Nature communications} \textbf{\bibinfo{volume}{5}}
  (\bibinfo{year}{2014}).

\bibitem[{\citenamefont{White et~al.}(1989)\citenamefont{White, Scalapino,
  Sugar, Loh, Gubernatis, and Scalettar}}]{White_PRB_1989}
\bibinfo{author}{\bibfnamefont{S.~R.} \bibnamefont{White}},
  \bibinfo{author}{\bibfnamefont{D.~J.} \bibnamefont{Scalapino}},
  \bibinfo{author}{\bibfnamefont{R.~L.} \bibnamefont{Sugar}},
  \bibinfo{author}{\bibfnamefont{E.~Y.} \bibnamefont{Loh}},
  \bibinfo{author}{\bibfnamefont{J.~E.} \bibnamefont{Gubernatis}},
  \bibnamefont{and} \bibinfo{author}{\bibfnamefont{R.~T.}
  \bibnamefont{Scalettar}}, \bibinfo{journal}{Phys. Rev. B}
  \textbf{\bibinfo{volume}{40}}, \bibinfo{pages}{506} (\bibinfo{year}{1989}).

\bibitem[{\citenamefont{Wang et~al.}(2011)\citenamefont{Wang, de' Medici, and
  Millis}}]{Millis_PRB_2011}
\bibinfo{author}{\bibfnamefont{X.}~\bibnamefont{Wang}},
  \bibinfo{author}{\bibfnamefont{L.}~\bibnamefont{de' Medici}},
  \bibnamefont{and} \bibinfo{author}{\bibfnamefont{A.~J.}
  \bibnamefont{Millis}}, \bibinfo{journal}{Phys. Rev. B}
  \textbf{\bibinfo{volume}{83}}, \bibinfo{pages}{094501}
  (\bibinfo{year}{2011}).

\bibitem[{\citenamefont{Kiesel and Thomale}(2012)}]{Kiesel_PRB_2012}
\bibinfo{author}{\bibfnamefont{M.~L.} \bibnamefont{Kiesel}} \bibnamefont{and}
  \bibinfo{author}{\bibfnamefont{R.}~\bibnamefont{Thomale}},
  \bibinfo{journal}{Phys. Rev. B} \textbf{\bibinfo{volume}{86}},
  \bibinfo{pages}{121105} (\bibinfo{year}{2012}).

\bibitem[{\citenamefont{Kiesel et~al.}(2013)\citenamefont{Kiesel, Platt, and
  Thomale}}]{Kiesel_PRL_2013}
\bibinfo{author}{\bibfnamefont{M.~L.} \bibnamefont{Kiesel}},
  \bibinfo{author}{\bibfnamefont{C.}~\bibnamefont{Platt}}, \bibnamefont{and}
  \bibinfo{author}{\bibfnamefont{R.}~\bibnamefont{Thomale}},
  \bibinfo{journal}{Phys. Rev. Lett.} \textbf{\bibinfo{volume}{110}},
  \bibinfo{pages}{126405} (\bibinfo{year}{2013}).

\bibitem[{\citenamefont{Bauer et~al.}(2013)\citenamefont{Bauer, Keller, Dolfi,
  Trebst, and Ludwig}}]{Ludwig_arXiv_2014}
\bibinfo{author}{\bibfnamefont{B.}~\bibnamefont{Bauer}},
  \bibinfo{author}{\bibfnamefont{B.~P.} \bibnamefont{Keller}},
  \bibinfo{author}{\bibfnamefont{M.}~\bibnamefont{Dolfi}},
  \bibinfo{author}{\bibfnamefont{S.}~\bibnamefont{Trebst}}, \bibnamefont{and}
  \bibinfo{author}{\bibfnamefont{A.~W.~W.} \bibnamefont{Ludwig}},
  \bibinfo{journal}{arXiv:1303.6963}  (\bibinfo{year}{2013}).

\bibitem[{\citenamefont{Di~Matteo and Norman}(2012)}]{Matteo_PRB_2012}
\bibinfo{author}{\bibfnamefont{S.}~\bibnamefont{Di~Matteo}} \bibnamefont{and}
  \bibinfo{author}{\bibfnamefont{M.~R.} \bibnamefont{Norman}},
  \bibinfo{journal}{Phys. Rev. B} \textbf{\bibinfo{volume}{85}},
  \bibinfo{pages}{235143} (\bibinfo{year}{2012}).

\bibitem[{\citenamefont{Varma}(2013)}]{Varma_arXiv_2013}
\bibinfo{author}{\bibfnamefont{C.~M.} \bibnamefont{Varma}},
  \bibinfo{journal}{arXiv:1307.1494}  (\bibinfo{year}{2013}).

\bibitem[{\citenamefont{Yakovenko}(2014)}]{Yakovenko_arXiv_2014}
\bibinfo{author}{\bibfnamefont{V.~M.} \bibnamefont{Yakovenko}},
  \bibinfo{journal}{arXiv:1409.2183}  (\bibinfo{year}{2014}).

\bibitem[{\citenamefont{Scagnoli et~al.}(2011)\citenamefont{Scagnoli, Staub,
  Bodenthin, de~Souza, Garc\'{i}a-Fern\'{a}ndez, Garganourakis, Boothroyd,
  Prabhakaran, and Lovesey}}]{Scagnoli_Science_2011}
\bibinfo{author}{\bibfnamefont{V.}~\bibnamefont{Scagnoli}},
  \bibinfo{author}{\bibfnamefont{U.}~\bibnamefont{Staub}},
  \bibinfo{author}{\bibfnamefont{Y.}~\bibnamefont{Bodenthin}},
  \bibinfo{author}{\bibfnamefont{R.~A.} \bibnamefont{de~Souza}},
  \bibinfo{author}{\bibfnamefont{M.}~\bibnamefont{Garc\'{i}a-Fern\'{a}ndez}},
  \bibinfo{author}{\bibfnamefont{M.}~\bibnamefont{Garganourakis}},
  \bibinfo{author}{\bibfnamefont{A.~T.} \bibnamefont{Boothroyd}},
  \bibinfo{author}{\bibfnamefont{D.}~\bibnamefont{Prabhakaran}},
  \bibnamefont{and} \bibinfo{author}{\bibfnamefont{S.~W.}
  \bibnamefont{Lovesey}}, \bibinfo{journal}{Science}
  \textbf{\bibinfo{volume}{322}}, \bibinfo{pages}{696} (\bibinfo{year}{2011}).

\bibitem[{\citenamefont{Zhu et~al.}(2013)\citenamefont{Zhu, Aji, and
  Varma}}]{Varma_PRB_2013}
\bibinfo{author}{\bibfnamefont{L.}~\bibnamefont{Zhu}},
  \bibinfo{author}{\bibfnamefont{V.}~\bibnamefont{Aji}}, \bibnamefont{and}
  \bibinfo{author}{\bibfnamefont{C.~M.} \bibnamefont{Varma}},
  \bibinfo{journal}{Phys. Rev. B} \textbf{\bibinfo{volume}{87}},
  \bibinfo{pages}{035427} (\bibinfo{year}{2013}).

\end{thebibliography}

\end{document}